\documentclass[aps,prb,preprint,superscriptaddress,reprint,preprintnumbers,amssymb,longbibliography]{revtex4-1}

\usepackage{amsmath}
\usepackage{amssymb}
\usepackage{braket}
\usepackage{bbold}
\usepackage[version=3]{mhchem}
\usepackage{graphicx}
\usepackage{color}
\usepackage{units}
\usepackage{pifont}
 \usepackage[caption=false]{subfig}
\usepackage{multirow}
\usepackage[dvipsnames]{xcolor}
\usepackage{hyperref}
\usepackage[final]{changes}

\DeclareMathOperator{\Tr}{Tr}

\AtBeginDocument{\usepackage{booktabs}}

\definechangesauthor[name=Stephan Mohr, color=red]{SM}
\definechangesauthor[name={Stephan Mohr}, color=ForestGreen]{LG}
\definechangesauthor[name={Stephan Mohr}, color=blue]{LR}
\definechangesauthor[name={Stephan Mohr}, color=magenta]{MM}
\definechangesauthor[name={done}, color=black]{done}

\newcommand{\good}{\textcolor{ForestGreen}{\ding{52}}}
\newcommand{\bad}{\textcolor{red}{\ding{56}}}

\newcommand{\tablesize}{\scriptsize}

\newcommand{\hpfig}[1]{\includegraphics[width=0.443\textwidth]{#1}}
\newcommand{\qpfig}[1]{\includegraphics[width=0.209\textwidth]{#1}}
\newcommand{\dosao}{\hpfig{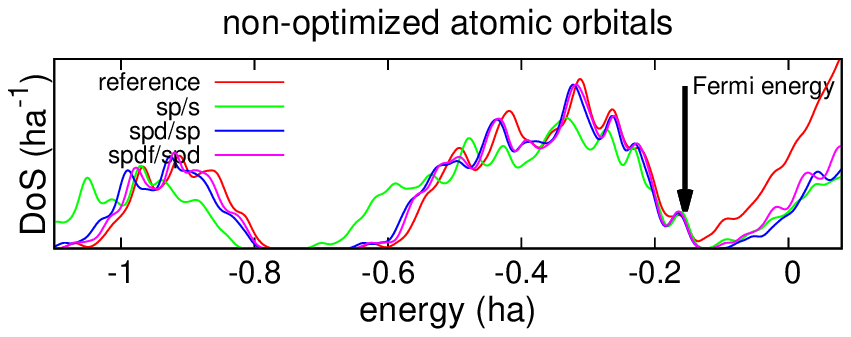}}
\newcommand{\dossf}{\hpfig{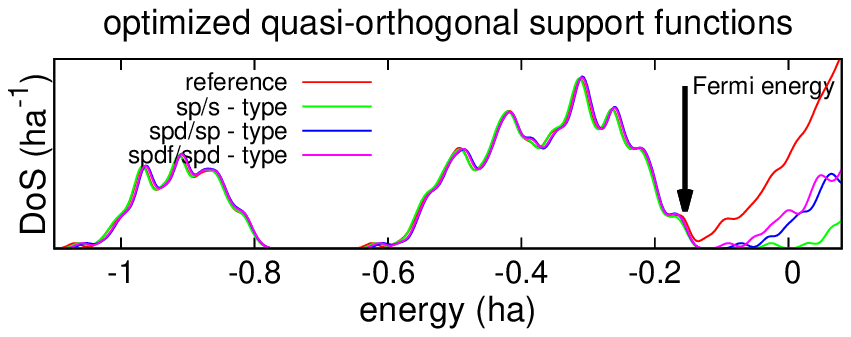}}
\newcommand{\pisfm}{\qpfig{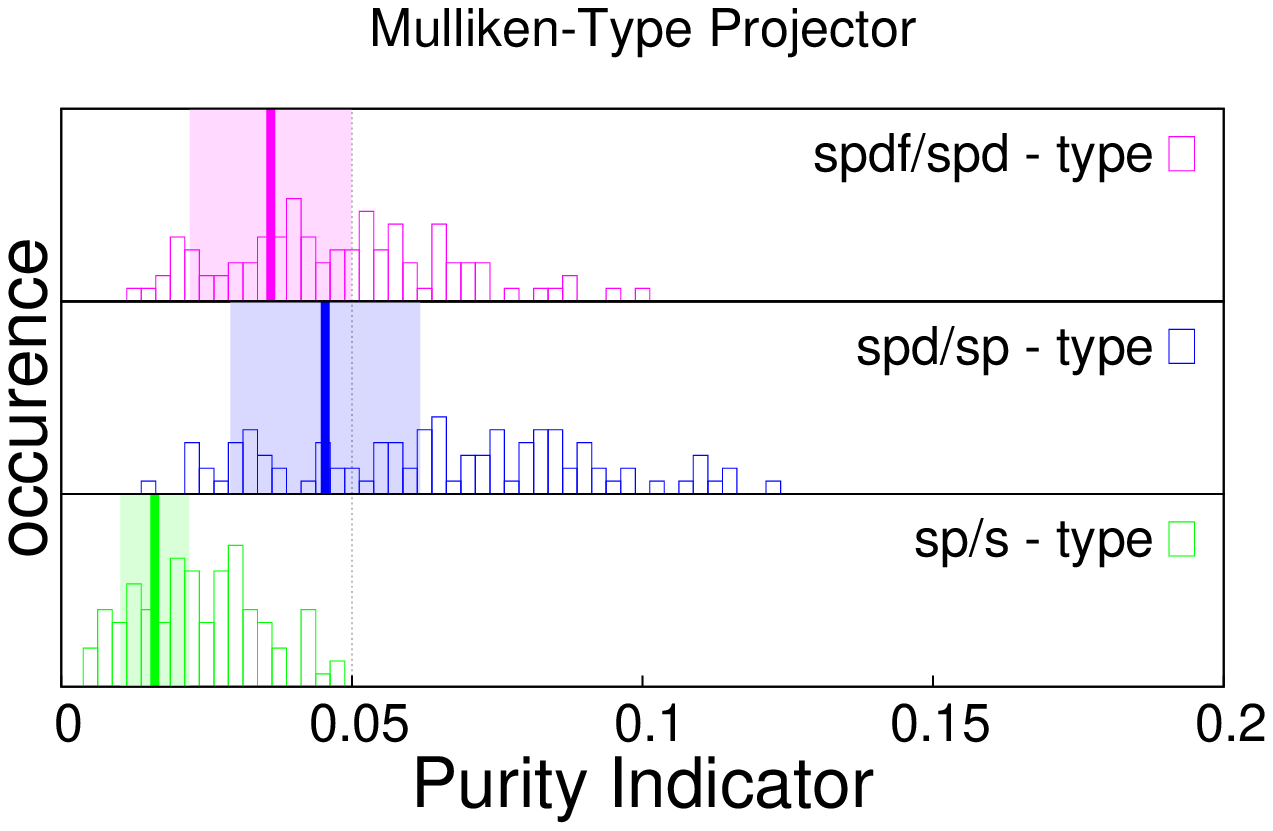}}
\newcommand{\pisfl}{\qpfig{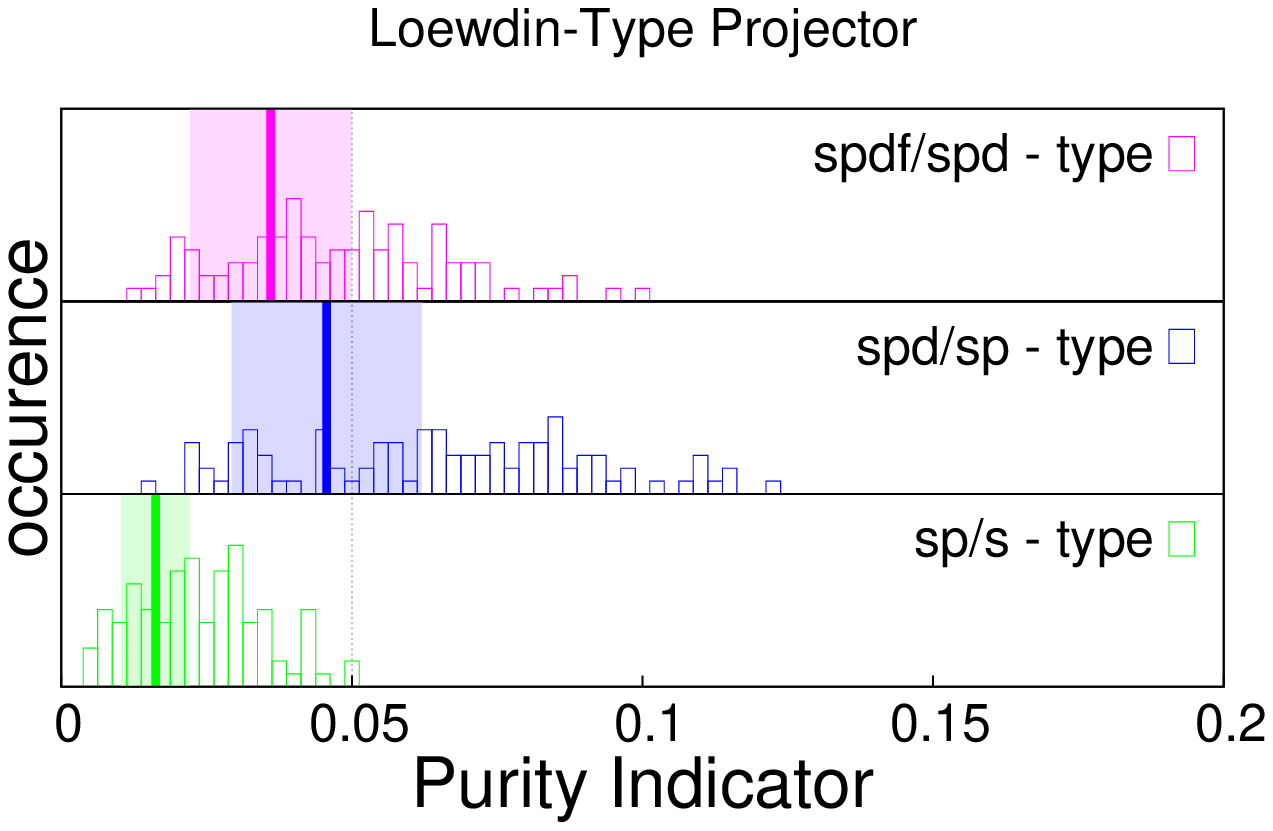}}
\newcommand{\piaom}{\qpfig{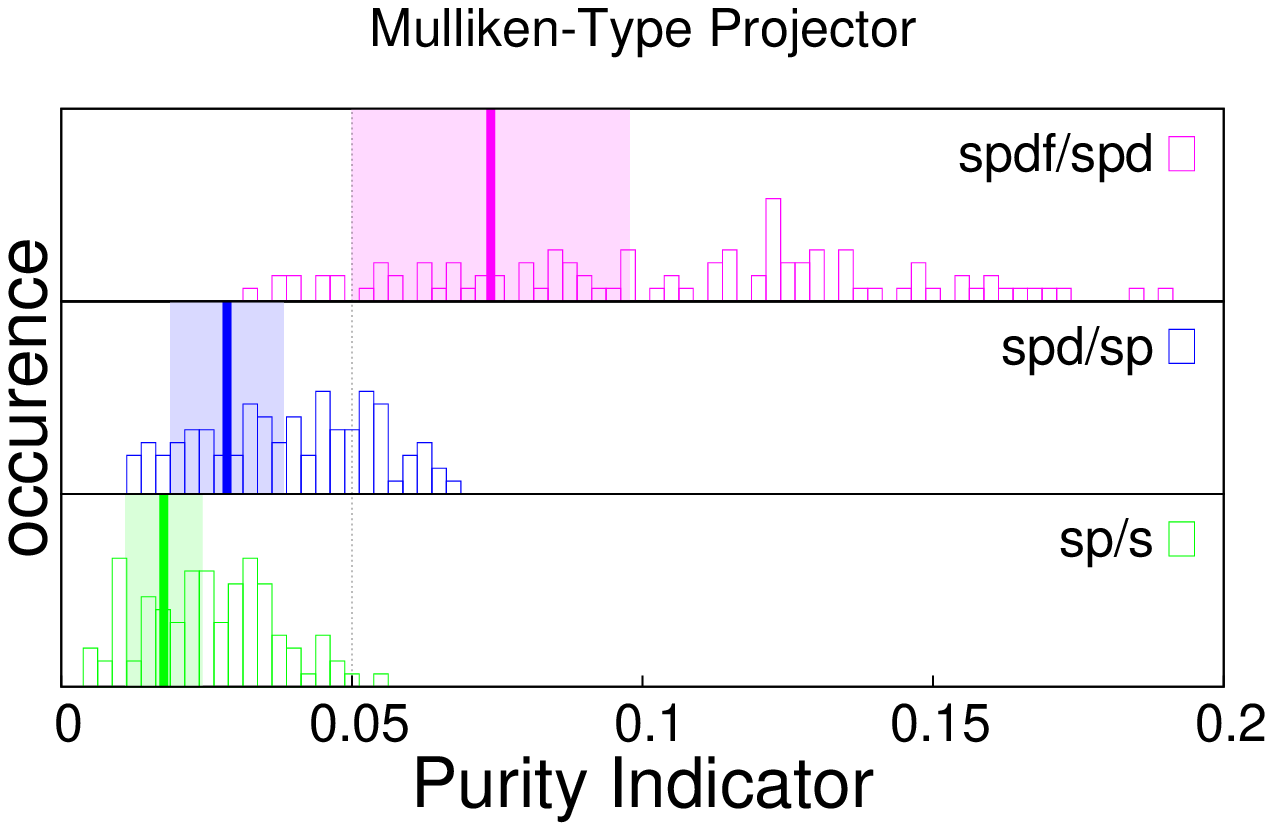}}
\newcommand{\piaol}{\qpfig{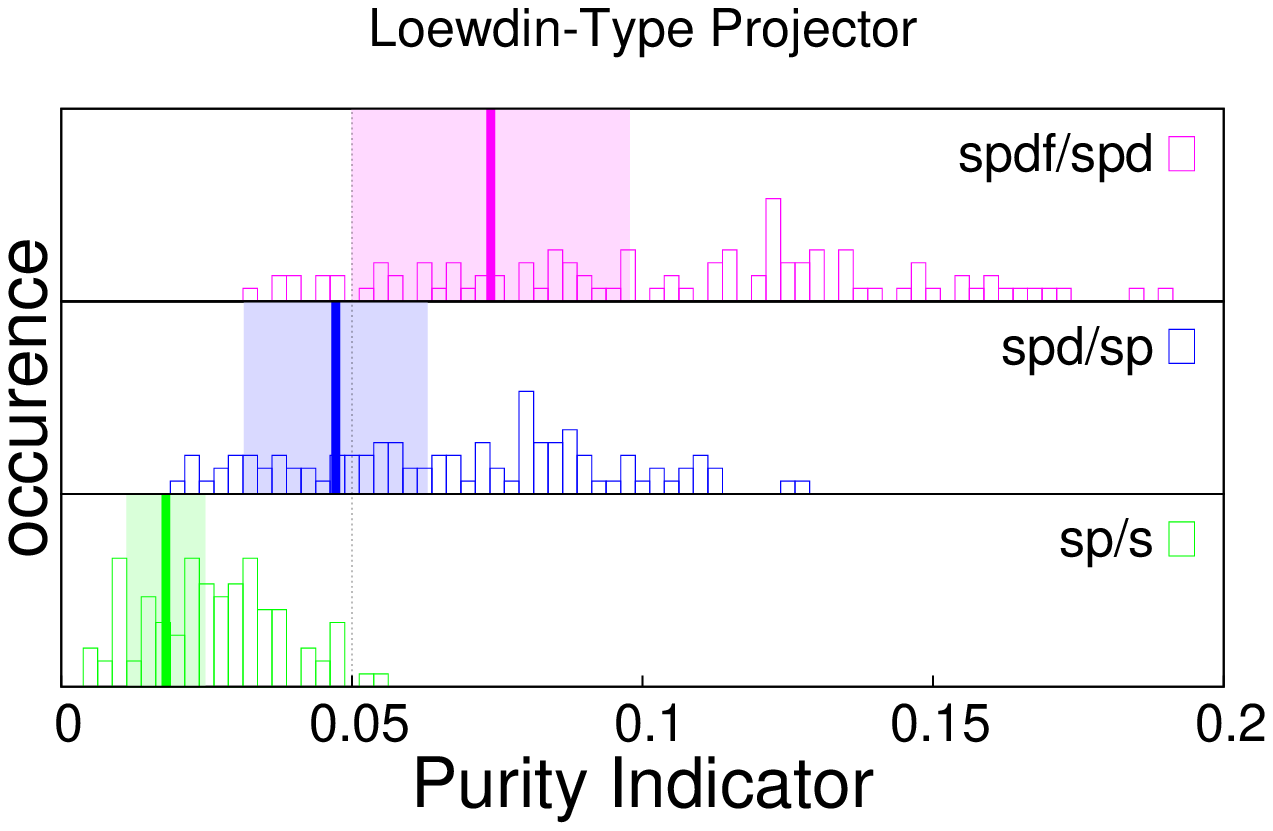}}
\newcommand{\dsfm}{\qpfig{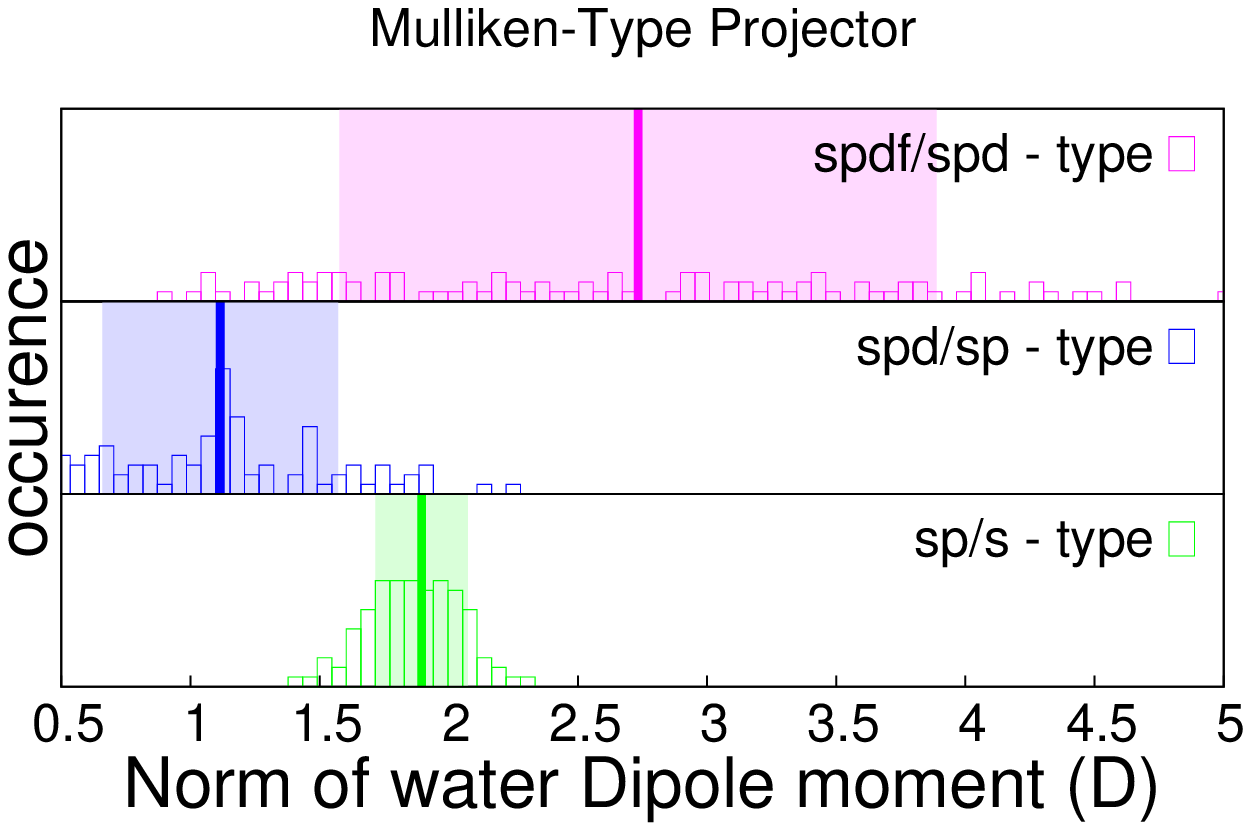}}
\newcommand{\dsfl}{\qpfig{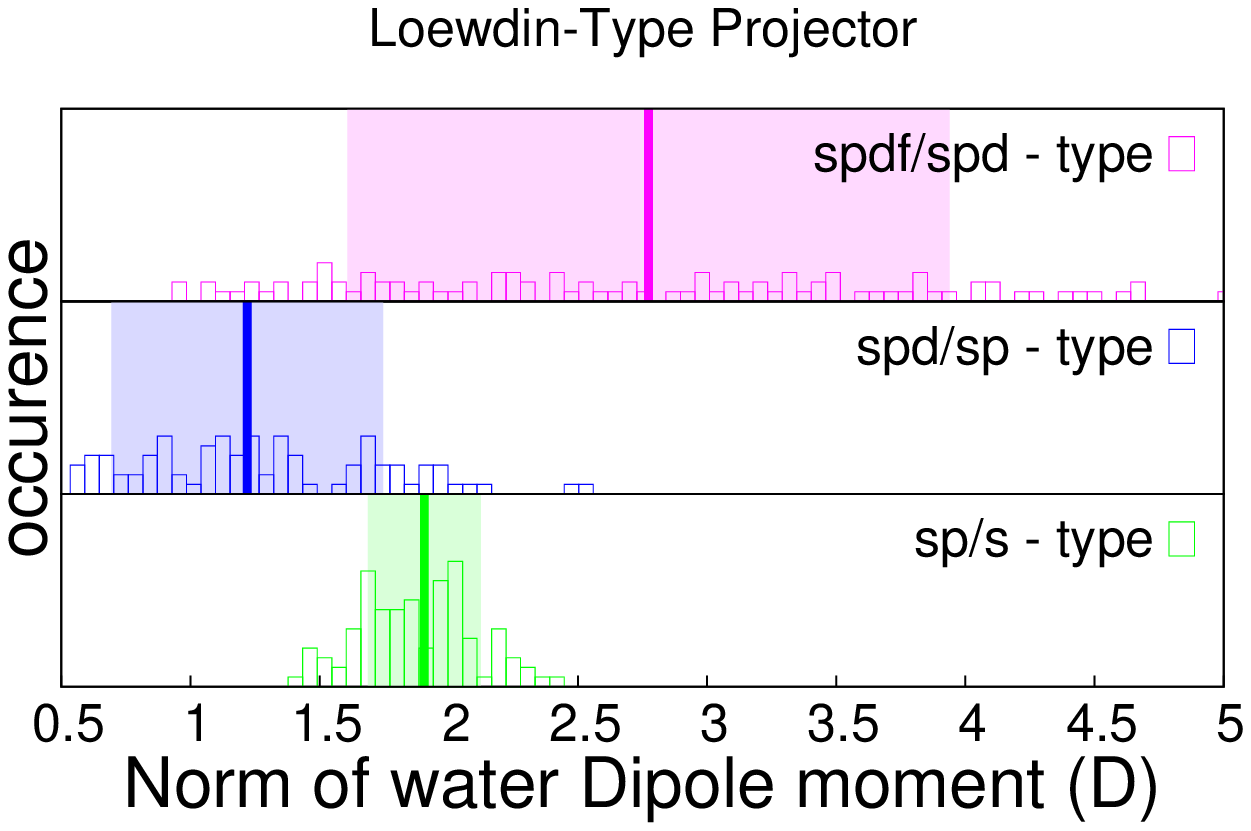}}
\newcommand{\daom}{\qpfig{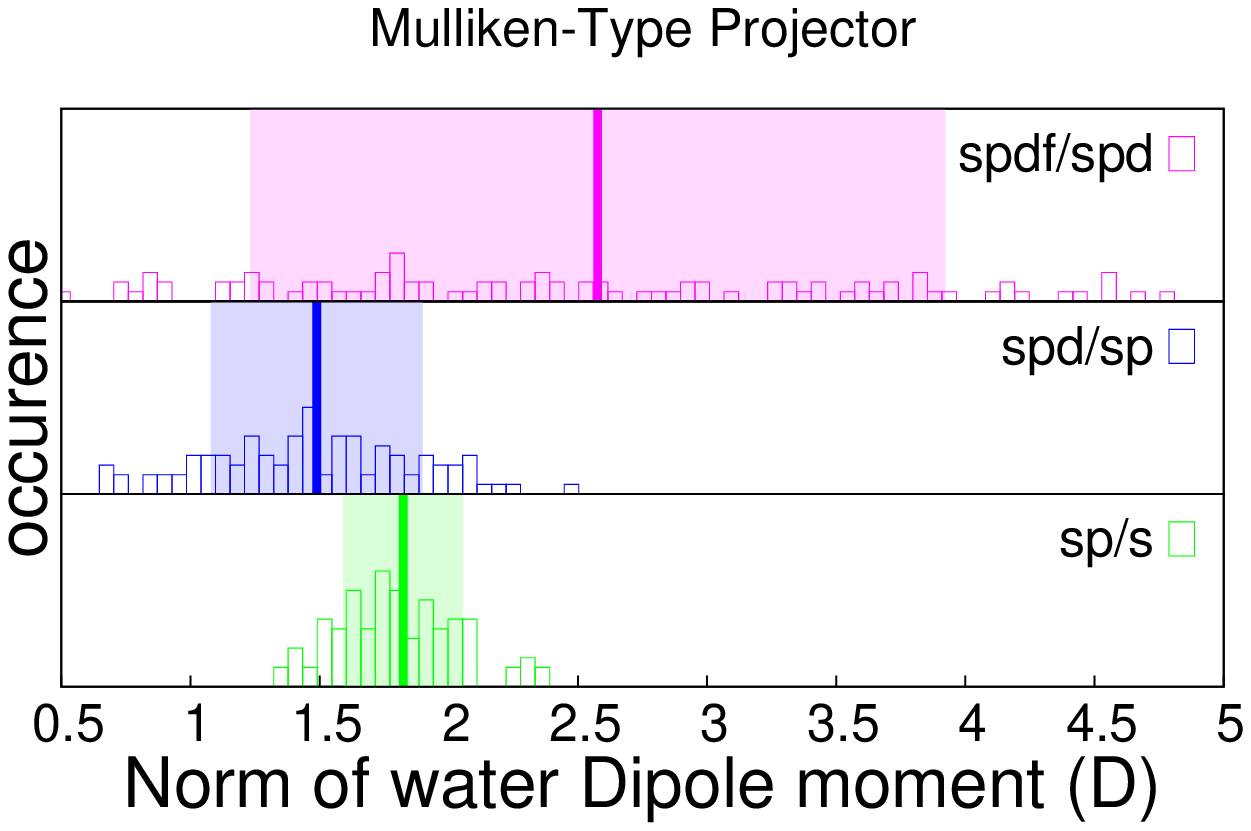}}
\newcommand{\daol}{\qpfig{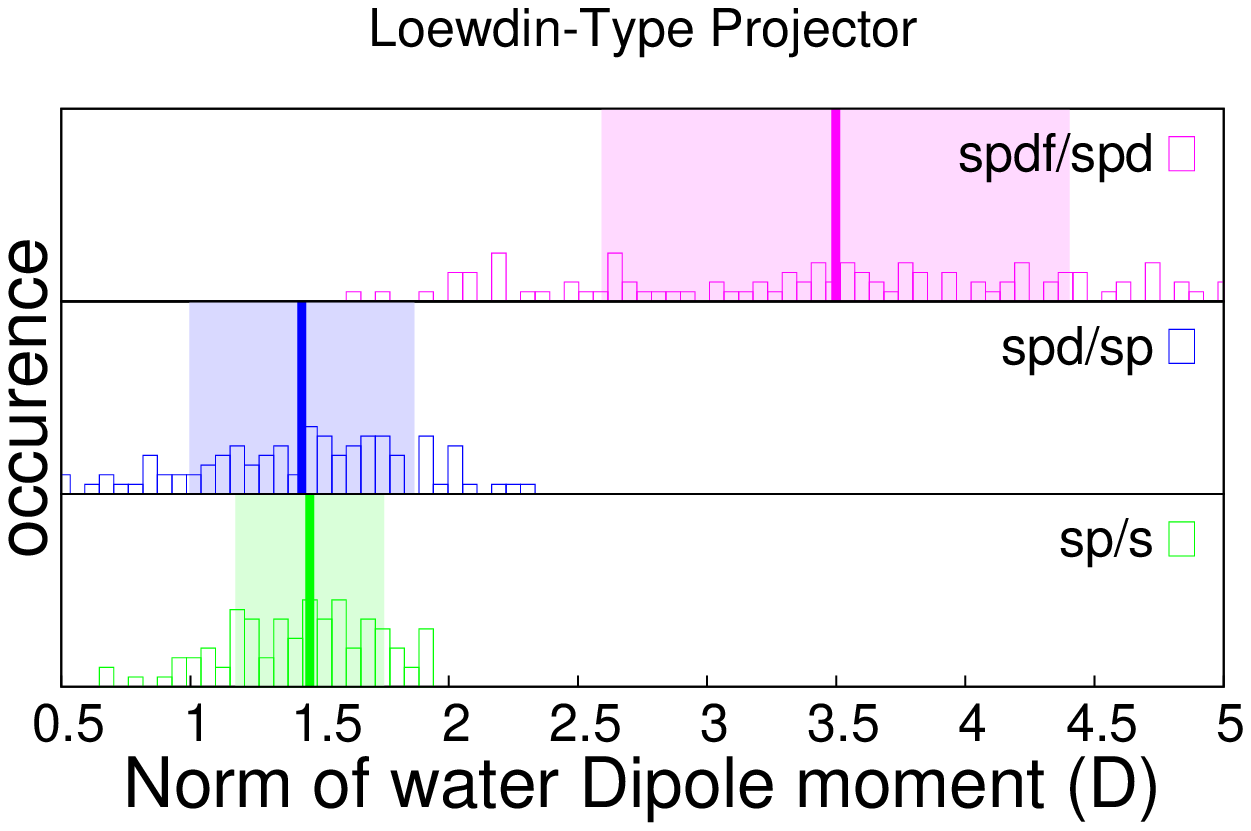}}
\newcommand{\rulesep}{\unskip\ \vrule}

\begin{document}


\author{Stephan Mohr}
\affiliation{Barcelona Supercomputing Center (BSC)}
\email{stephan.mohr@bsc.es}

\author{Michel Masella}
\affiliation{Laboratoire de Biologie Structurale et Radiologie, Service de Bio\'energ\'etique, Biologie Structurale et M\'ecanisme,Institut de Biologie et de Technologie de Saclay, CEA Saclay, F-91191 Gif-sur-Yvette Cedex, France}

\author{Laura E. Ratcliff}
\affiliation{Argonne Leadership Computing Facility, Argonne National Laboratory, Illinois 60439, USA}

\author{Luigi Genovese}
\affiliation{Univ.\ Grenoble Alpes, INAC-MEM, L\_Sim, F-38000 Grenoble, France}
\affiliation{CEA, INAC-MEM, L\_Sim, F-38000 Grenoble, France}
\email{luigi.genovese@cea.fr}

\title{Complexity Reduction in Large Quantum Systems: Fragment Identification and Population Analysis via a Local Optimized Minimal Basis}


\date{\today}

\begin{abstract}
We present, within Kohn-Sham Density Functional Theory calculations, a quantitative method to identify and assess the partitioning of a large quantum mechanical system into fragments.
We then show how within this framework
simple generalizations of other well-known population analyses can be used 
to extract, from first principles, reliable electrostatic multipoles for the identified fragments.
Our approach reduces arbitrariness in the fragmentation procedure, and enables the possibility to assess, quantitatively, whether the corresponding fragment multipoles can be interpreted as observable quantities associated to a system's moiety.
By applying our formalism within the code \textsc{BigDFT}, we show that
the use of a minimal set of in-situ optimized basis functions allows at the same time a proper fragment definition and an accurate description of the electronic structure.
\end{abstract}

 
\maketitle

\section{Introduction}

First-principles computational quantum mechanical (QM) approaches are nowadays able to provide reasonably accurate modelizations for a wide variety of systems. In particular  Density Functional Theory (DFT) approaches based on the Kohn-Sham (KS) formalism~\cite{hohenberg-inhomogeneous-1964,kohn-self_consistent-1965} are probably the most popular, usually presenting a good compromise between accuracy and computational complexity. Nonetheless, even when a DFT approach gives an accurate description of a microscopic system, it is advantageous in certain situations to consider an effective complexity reduction (ECR), allowing one 
to get the same level of accuracy by explicitly considering fewer degrees of freedom.

The fundamental principle of an ECR lies in the \emph{identification} of the essential moieties (i.e.\ ``fragments'') of a system out of an atomic description.
These fragments should then, in turn, be treated with an adequate methodology depending on the specific needs.
A less complex description of a system might contribute to decreasing the computational cost of the calculation; however, this is not the sole advantage of an ECR.
Within such a scheme the observable quantities that can be extracted for the system as a whole can also be assigned \emph{separately} to each of the fragments. 
Such a procedure allows a better understanding of the relevant mechanisms which govern the interactions among the constituents of the system, together with the design and validation of coarse-graining models, that are adapted for systems of length scales for which atomistic QM models would be unnecessarily costly or even out of reach~\cite{ratcliff-2016-challenges}.

A great variety of fragmentation methods has been developed;
an exhaustive overview can be found in Refs.~\citenum{gordon-fragmentation-2012,collins-energy-based-2015}.
In all of these methods the fragments are chosen based on pre-defined conditions, such as geometrical criteria or basic chemical intuition, and there is no possibility for verifying \emph{a posteriori} whether the chosen fragmentation is sensible or not for the actual setup of the simulation.
A typical observable which is then determined for each moiety is the electronic charge, extracted from the charge density of the QM calculation of the entire system, partitioned among the fragments.
Typically, the attention is focused on the atoms composing the system,
and a large number of \emph{atomic} charge population analyses have been developed.

All of these population analyses have their advantages and shortcomings, and applied to the very same system they might even give considerably different results~\cite{wiberg-comparison-1993,fonseca-voronoi-2004}.
However, from a conceptual point of view,
all of them suffer from the same problem: the electrostatic multipoles of the atoms, considered separately, are \emph{not} observable quantities of a QM system.
The only electrostatic quantities that are truly QM observables are the charge multipoles of the whole system, which are of course well defined and independent of the basis set as they are a function of the charge density of the system, which should not alter under changes of the basis; all the methods should yield the \emph{same} values, provided of course an adequate level of completeness. 
For a portion of the system like an atom or a fragment, electrostatic multipoles become ``pseudo-observables'', whose pertinence depends on the method chosen, the basis set used, and the definition of the subsystem itself.
In the context of ECR methods based on electrostatic multipoles of a subsystem, this is a crucial fact that has to be taken into account.
In other words, such methods suffer from two (somehow related) shortcomings: Firstly it is not possible to systematically validate the pertinence of the chosen fragmentation scheme, and secondly they do not allow one to quantify whether the electrostatic fragment multipoles extracted from the QM simulation can be considered as physical ``pseudo-observables'', i.e.\ with a meaningful physico-chemical interpretation. 

In this paper we propose a general theoretical scheme to identify subsystems (i.e.\ fragments)
out of a large QM system, accounting for the aforementioned problems.
Our method, which we will denote as ``purity indicator'', allows one to assess quantitatively the suitability of the employed basis set for the chosen population method;
thanks to this information
we can therefore verify, in a quantitative way, whether a given fragmentation of a QM system is compatible with the employed \emph{combination} of 
the basis set and the population method. 
We show that in situations where this is the case, the electrostatic quantities calculated on the pre-defined fragment moieties have
the reliability of QM observables and can be interpreted as such.
On the other hand, the same technique might also be employed to determine \emph{a posteriori}, i.e.\ based on a QM calculation of the entire system, which are the 
essential moieties that can be considered as well defined entities for the actual fragmentation method and basis set. 

Our approach is based on the \emph{density matrix} of the system, which is a well-defined QM entity; this is in contrast to other popular QM-based fragmentation schemes, such as the Fragment Molecular Orbital (FMO) approach~\cite{choi-reducing-2012,kitaura-fragment-1999,fedorov-extending-2007} or X-Pol~\cite{gao-toward-1997,gao-a-molecular-1998,wierzchowski-hydrogen-2003,xie-design-2007,xie-the-variational-2008,xie-incorporation-2008}, where only the pre-selected fragments are treated on a stringent ab-initio level.
Like all methods based on the density matrix, this intrinsically only gives access to integrated quantities such as the charge or the total energy.
This is in contrast to methods that explicitly calculate the wave functions within a fragmentation approach, which have also the advantage that they can be applied to excited states~\cite{wu-linear-scaling-2011,liu-photoexcitation-2014,li-localization-2014,li-localization-2017}.
Within our framework we further point out the competitive advantages of a \textit{minimal and optimized} basis set in the context of ECR methods.
Using the purity indicator it can be shown that such a computational setup considerably simplifies both the fragment identification and multipole assignment.
We additionally demonstrate that within this setup, even straightforward generalizations of pioneering approaches like Mulliken and L\"owdin population analyses provide high quality and chemically sound results, whose reliability can be assessed in a quantitative way.

The outline of the paper is as follows. We first present in Sec.~\ref{sec: Methodology} the basic ideas of the identification and assessment of the system fragments and the calculation of the associated multipoles. 
In Sec.~\ref{sec: fragment and basis} we then discuss the important relation between the fragment definition and the employed basis set, by defining \textit{a priori} a specific fragment definition and population scheme and then searching for the optimal basis for this setup.
In Sec.~\ref{sec:DNA in water (results)} we then inverse the focus and identify --- within the setup of a given basis --- the fragments for a large complex molecule in solvation.

\section{Methodology}
\label{sec: Methodology}

\label{sec:Theory}

\subsection{Fragment identification and assessment} \label{sec:identification}

Let us assume that a QM system can be split into $M$ different fragments.
This means that, in a ``QM sense'', the wave function can be 
approximated by a \emph{separable} wave function, i.e.\
\begin{equation}
 |\Psi\rangle \simeq |\Psi^1\rangle \otimes |\Psi^2\rangle \otimes \cdots \otimes |\Psi^M\rangle \; ,
 \label{eq:wave function decomposition}
\end{equation}
where each of the states $|\Psi^\mathfrak F\rangle$ is associated to the quantum description of the fragment $\mathfrak F$.

The simplest case where the above assumption is valid is the \emph{cluster decomposition}, 
which also implies (the opposite is not necessarily true) that a spatial separation can be readily defined between the system elements and their respective wave functions do not overlap. In addition the Hilbert spaces of the different subsystems can be factorized in different subspaces where QM observables are correctly defined.
To define pseudo-observables like the electrostatic multipoles of a system element,
we are interested in a suitable realization of the above situation for a KS-DFT computation.

Let us suppose that we can express the (one-body) density operator of the system 
in a finite set of  localized basis functions $|\phi_\alpha\rangle$ as follows:
\begin{equation}
 \hat F =\sum_{\alpha,\beta} |\phi_\alpha\rangle K_{\alpha\beta}\langle \phi_\beta| \;.
 \label{eq:density_matrix_in_support_function_basis}
\end{equation}
This is a common ansatz for large scale DFT calculations~\cite{hernandez-self-1995,skylaris-introducing-2005,bowler-recent-2006}.
In the following, the basis functions $|\phi_\alpha\rangle$ will be called \emph{support functions},
and the matrix $\mathbf K$ will denote the \emph{kernel}. If $\hat F$ is obtained from  a many-body wave function $|\Psi\rangle$ expressed via a single Slater determinant the above density matrix is idempotent, i.e.\ $\hat F^2 = \hat F$, and the kernel is pure, i.e.\ it obeys \ $\mathbf{K S K} = \mathbf K$, where $S_{\alpha\beta} = \langle \phi_\alpha | \phi_\beta \rangle$ is the overlap matrix among the support functions.

When a QM system is genuinely separable, it should be possible to define a projector operator $\hat W^{\mathfrak F}$  associated with each fragment $\mathfrak F$ 
such that $\hat W^{\mathfrak F}|\Psi\rangle = | \Psi^\mathfrak F\rangle $.
For such a separable system, the QM measure of any observable $\hat O$ may also be associated with the fragment, by evaluating $\Tr\left({\hat F \hat W^\mathfrak F \hat O}\right)$%
. The quantity
$\hat F^\mathfrak{F} = \hat F \hat W^\mathfrak F$ may thus be referred to as the ``fragment density matrix''. 
For a separable system such a density operator is idempotent by construction.
Separability of the associated many-body wavefunctions $|\Psi^\mathfrak F\rangle$ also implies that different fragments are orthogonal, i.e.
$\hat F^\mathfrak{F} \hat F^\mathfrak{G} =\hat F^\mathfrak{F} \delta_{\mathfrak{F G}}$.
For a reasonable fragment definition we should require that the complete set of fragments 
represents a partitioning of the system, i.e.
\begin{equation}\label{partition}
 \sum_\mathfrak{F} \hat F^\mathfrak{F} = \hat F \;.
\end{equation}

To proceed further we assume that $\hat W^\mathfrak{F}$ can be provided in the same basis set as that used to describe the density matrix:
\begin{equation}
 \hat W^\mathfrak{F} = \sum_{\mu,\nu}\ket{\phi_\mu}R_{\mu\nu}^\mathfrak{F}\bra{\phi_\nu} \;,
 \label{eq:projector_ansatz}
\end{equation}
where the (still to be defined) matrix $\mathbf R^\mathfrak F$ determines the character of the fragment projection operator;
several examples will be given later.

For a QM system that is not genuinely separable, a ``fragment quantity'' is \emph{not} a well-defined quantum observable.
Of course there is no universal recipe to define the fragment partitioning,
which leads to the question of the \emph{pertinence} of the operator $\hat W^\mathfrak F$.
We would like then to \emph{quantify} the reliability of the identification of  $\mathfrak{F}$ as a system's moiety by the projector defined from $\mathbf R^\mathfrak F$.
If such a fragment restriction makes sense, the operator $\hat F^\mathfrak{F}\equiv \hat F \hat W^\mathfrak{F}$ should --- following the above discussion --- be idempotent, i.e.\ $\left(\hat F^\mathfrak{F}\right)^2 = \hat F^\mathfrak{F}$.
Hence, the quantity
\begin{equation}\label{idempotency}
 \Tr\left(\left(\hat F^\mathfrak{F}\right)^2 - \hat F^\mathfrak{F}\right) = \Tr\left( \left(\mathbf{ K S}^\mathfrak{F} \right)^2 - \mathbf{ K S}^\mathfrak{F}\right) \;, 
\end{equation}
with $\mathbf S^\mathfrak{F}\equiv \mathbf S \mathbf R^\mathfrak{F} \mathbf S$, is well suited
to quantify the pertinence of fragment $\mathfrak{F}$ being considered as a genuine fragment of the full system.
We will call this quantity from now on the \emph{purity indicator};
the closer this index is to zero the more properly the fragment $\mathfrak F$ is identified.
In order to define an intensive quantity we may additionally normalize the purity indicator and thus consider the quantity
\begin{equation}
 \Pi = \frac{1}{q} \Tr\left( \left(\mathbf{ K S}^\mathfrak{F} \right)^2 - \mathbf{ K S}^\mathfrak{F}\right) \;,
 \label{eq:normalized_purity_indicator}
\end{equation}
where we indicate with $q$ the total number of electrons of the fragment in gas phase.

The above derivation makes apparent that the purity indicator is an explicit functional of the matrix $\mathbf R^\mathfrak F$ and the basis set $\{\phi_\mu\}$.
Consequently it is evident that this quantity is \emph{not} a QM observable.
Rather, it has to be interpreted as a \emph{necessary} condition for the matrix $\mathbf R^\mathfrak F$ to be meaningful for the identification of a fragment within a given basis. If this condition is not fulfilled and the purity indicator is high,
it is unlikely that the value of $\Tr\left({\hat F^\mathfrak F \hat O}\right)$ can be associated with an observable quantity of the fragment $\mathfrak F$.

It is important to stress here that these criteria are less stringent than a simple spatial separation between the fragments, as they are defined in terms of entries of the density matrix operator in the employed basis set.
As an illustrative example for a proper fragmentation, we can choose any operator that selects one (or more) KS orbitals,
\begin{equation}\label{wj}
 \hat W^j= \ket{\psi_j}\bra{\psi_j} \;.
\end{equation}
Indeed this is a suitable definition:
Due to the orthonormality of the KS orbitals the trace in Eq.~\eqref{idempotency} is exactly zero, and $\sum_j \hat W^j = \hat F$, 
thus also fulfilling Eq.~\eqref{partition}. 
This is consistent with the obvious consideration that it makes sense to project density matrix-related quantities onto a subset of
the KS orbitals.

\subsection{Atomic charge population analyses}
\label{sec: Fragment Population Analysis}
Traditionally the most common choice for the fragments are the individual atoms.
We therefore want to briefly revisit some popular atomic charge population analyses.
A pioneering example
is provided by  the Mulliken approach~\cite{mulliken-electronic-1955}, which directly uses the atomically localized basis functions in which the QM molecular orbitals are expressed, and is thus conceptually very simple.
On the other hand the outcome of the Mulliken analysis depends strongly on the used basis set (see e.g.\ Refs.~\citenum{reed-natural-1985,politzer-comparison-1971} and references therein) and a bad choice might yield completely misleading results.
The L\"owdin population analysis~\cite{loewdin-on-1950,loewdin-on-1970} is akin in spirit, with the difference that it works with a set of orthonormalized orbitals. 
The strong sensitivity with respect to the basis set is considerably reduced by an approach like the natural population analysis (NPA)~\cite{reed-natural-1985}, which evaluates the atomic charges as the occupancies of a set of special ``Natural Atomic Orbitals'' (NAO).
The advantage of NPA over Mulliken and L\"owdin is that the first one is built upon ``wavefunction-based'' physical concepts, like the definition of the Natural Atomic Orbitals, whereas the latter ones rely on a partitioning scheme that considers \emph{all} the basis functions on the atom on an equal footing.

Now we want to see how this connects to our general framework,
by applying it to KS-DFT calculations and comparing with the aforementioned well-established methods.
If a fragment is a well defined and independent subsystem, there exists a set of ``fragment states'' $\ket{\psi_\mu^\mathfrak F}$ (which are eigenfunctions of the projector $\hat{W}^\mathfrak{F}$), together with their dual functions $\bra{\tilde{\psi}_\mu^\mathfrak F}$, thus fulfilling $\braket{\tilde{\psi}_\mu^\mathfrak F | \psi_\nu^\mathfrak F} = \delta_{\mu\nu}$. 
As we are here dealing with fragments formed by the individual atoms, we can in the same way assume atomic states $\ket{\psi_\mu^A}$ and define the projector $\hat W^A$ onto that atom 
by summing over them:
\begin{equation}
\hat W^A \equiv \sum_\mu |\psi_\mu^A \rangle \langle\tilde  \psi_\mu^A |\;.
\label{eq:projector_from_fragment_states}
\end{equation}

The most straightforward approach to identify fragments out of a system described by localized basis functions is to \emph{associate} a set of basis functions with a given atom $A$.
These atoms can then also eventually be combined to form a fragment $\mathfrak F$ constituted by this group of atoms, as will be discussed later.
The restriction to an atom $A$ can be implemented by the diagonal matrix $T^A_{\mu\nu}=\delta_{\mu\nu}\theta(A,\mu)$, 
where $\theta(A,\mu)$ is defined as
\begin{equation}
 \quad \theta(A,\mu) =   
  \begin{cases}
    1 & \mbox{if } \mu \text{ is associated with atom } A \\ 
    0 & \mbox{otherwise } 
    \end{cases} \;.
\end{equation}
Such an association is clearly arbitrary and is based on considerations about the (presumed) center of the associated basis function.
Information about the basis extensions are often neglected and might lead to unreliable partitionings, as the clear association of a basis function with an atom is not obvious any more.
When adopting this approach of fragment selection it is important to remember the previously mentioned bi-orthogonality and to distinguish between direct and dual ``fragment states''. 
Suppose we define $|\psi_\mu^A\rangle =\sum_{\beta} T^A_{\mu \beta}  |\phi_\beta\rangle$ as an atomic state.
The orthonormality constraint then imposes that $\langle \tilde \psi_\mu^A| = \sum_{\beta} \langle \phi_\beta| S^{-1}_{\beta \mu}$.
By plugging this into Eq.~\eqref{eq:projector_from_fragment_states} and comparing with Eq.~\eqref{eq:projector_ansatz} it follows that the projector matrix reads
\begin{equation}\label{mulli}
\mathbf{R}^A_M =  \mathbf{T}^A\mathbf{S}^{-1} \,.
\end{equation}
As will be shown later, such a definition corresponds to nothing other than the traditional Mulliken population analysis.

Proceeding in an analogous way, we can also define the fragment states in terms of a basis which is first orthogonalized, giving  $|\psi_\mu^A\rangle = \sum_{\beta \gamma} T^A_{\mu \beta} S^{-1/2}_{\beta \gamma} |\phi_\gamma\rangle$, 
and therefore 
$\langle \tilde \psi_\mu^A| = \langle \psi_\mu^A|$.
This leads to the projector matrix
\begin{equation}\label{loew}
\mathbf R^A_L = \mathbf S^{-1/2} \mathbf T^A \mathbf S^{-1/2} \;,
\end{equation}
which corresponds, as will be demonstrated later, to the definition of the L\"owdin population analysis. 

We may also revisit the NPA method under this light.
Here the degrees of freedom of the subsystem are defined in the basis of Natural Atomic Orbitals (NAO) which are by construction orthonormal.
These are generated in a procedure involving several steps, resulting in an expression that can be written as linear combinations (with coefficients $\mathbf B^A$) of the original basis functions projected on the atoms $A$ (see Ref.~\citenum{reed-natural-1985}):
\begin{equation}
|\psi^{A}_\mu\rangle = \sum_\beta B^A_{\mu\beta} |\phi_\beta\rangle\;.
\label{eq:NAO}
\end{equation}
Within this scheme the NAO projector operator is defined as
\begin{equation}
\mathbf R^A_{NAO} = {\mathbf B^A}^T \mathbf T^A \mathbf B^A  \;.
\label{eq:projector_matrix_NPA}
\end{equation}
The transformation matrix $\mathbf B^A$ is defined in such a way to ensure that the NAO are eigenstates of the density operator for a given atom, that can thus directly be written in terms of the sum over the NAO:
\begin{equation}
\hat F^A=\sum_\mu \theta(A,\mu)|\psi^{A}_\mu\rangle N^A_\mu \langle \psi^{A}_\mu| \;.
\end{equation}

The NPA method is considered to be more robust than the Mulliken and L\"owdin approaches, since it removes the strong dependence of the results on the basis set.
This superiority is related to the fact that basis sets with diffuse degrees of freedom often contain components that considerably contribute to the description of empty states.
In the NPA scheme their contribution is weighted by the 
eigenvalue $N_\alpha^A$, whereas in the Mulliken or L\"owdin scheme all the 
atomic components have the same weight.
A similar approach to NPA is the use of so-called AOIMs (atomic orbitals in molecular environments)~\cite{liu-a-method-1997}, which as well have the goal of providing a reliable and stable population analysis for variable (and in particular large) basis sets.
The AOIMs are defined as the solution of the single-electron Schr\"odinger equation with an effective potential given by the spherical average of the molecular potential centered on the given atom.
Once the AOIMs have been determined, a standard population scheme such as the Mulliken approach yields reliable and robust results.

For the Mulliken projector (Eq.~\eqref{mulli}), the condition of Eq.~\eqref{idempotency} corresponds to the idempotency of the matrix $\mathbf{K S T}^A$, i.e.\ the block of the $\mathbf{KS}$ matrix associated with the indices of atom $A$.
The L\"owdin projector (Eq.~\eqref{loew}), on the other hand, can be considered meaningful if the atomic block matrix of $\mathbf S^{1/2} \mathbf K \mathbf S^{1/2}$ 
is close to idempotency.
By orthogonality of the NAO, the NPA approach is idempotent if all the NPA eigenvalues $N^A_\alpha$ associated with the atom $A$ are 0 or 1.

A situation in which Mulliken and L\"owdin are unreliable corresponds to a setup that yields a non-pure atomic (or more general fragment) kernel.
The above considerations show that this non-purity is not only a consequence of a inappropriate fragment choice, but also related to the basis.
This is an important point, as it means that even simple population schemes might lead to unbiased and reliable results if the basis employed leads to pure fragment kernels.
Indeed we show later in Sec.~\ref{sec: fragment and basis} that, whenever it is possible to identify a sensible fragment, a \textit{minimal} basis leads --- for both Mulliken and L\"owdin --- to such a favorable situation.

\subsection{Generalized multipole decomposition}
\label{sec: multipole decomposition}
To analyze the features of the density matrix of a system, the most intuitive objects to use
 are the multipoles of the charge density $\rho(\mathbf r)$.
These read
\begin{multline} \label{basicqlm}
 Q^R_{\ell m} \equiv  \sqrt{\frac{4\pi}{2\ell + 1}} \int \mathcal{S}_{\ell m}(\mathbf r - \mathbf{r}_R) \rho(\mathbf r) \, \mathrm d \mathbf r \\ = 
 \sqrt{\frac{4\pi}{2\ell + 1}} \Tr \left( \hat F \hat{\mathcal{S}}^R_{\ell m} \right) =
 \Tr\left(\mathbf K \mathbf{P}^R_{\ell m} \right)
 \;,
\end{multline}
where we have defined the multipole matrices $\mathbf{P}^R_{\ell m}$ as
\begin{equation}
P^R_{\ell m;\alpha\beta} = \sqrt{\frac{4\pi}{2\ell+1}}\braket{\phi_\alpha|\hat {\mathcal{S}}^R_{\ell m}|\phi_\beta} \;.
 \label{eq:basic_equation_atomic_multipole_matrices}
\end{equation}
In the above equation the superscript $R$ indicates that the solid harmonic operators 
$\hat {\mathcal{S}}^R_{\ell m}(\mathbf r) \equiv \mathcal{S}_{\ell m}(\mathbf r - \mathbf{r}_R)$ are centered on the reference position $\mathbf r_R$;
their proper definitions are presented in Appendix~\ref{sec: Definition of the spherical harmonics and translation relations} for completeness.
We may therefore say that the electrostatic multipoles are functions of the density matrix 
and the center $\mathbf r_R$ of the reference system.

The resulting $Q^R_{\ell m}$ can however also be used for the calculation of multipoles with respect to a different origin $\mathbf r_{R'}$.
As is shown in more detail in Appendix~\ref{sec: Definition of the spherical harmonics and translation relations} we obtain the relation
\begin{equation}\label{shiftedmp}
 Q^{R'}_{\ell m} =  \sum_{\ell'=0}^\ell \sum_{m'=-\ell'}^{\ell'} Q^R_{\ell' m'} \mathcal C^{\ell m}_{\ell' m'}(\mathbf r_{R'}-\mathbf r_R)  \;,
\end{equation}
where the functions $\mathcal C^{\ell m}_{\ell' m'}(\mathbf r)$ can be
expressed in terms of the $\mathcal{S}_{\ell-\ell' m''}(\mathbf r)$.
For the important cases of the monopole and dipole components these equations are very simple and provide
\begin{subequations}
 \begin{align}
  Q^{R'}_{00} &= Q_{00}^R \;, \label{eq:molecular_dipole_from_atomic_multipoles_with_Slm_monopole} \\
  Q^{R'}_{1 m} &= \sqrt{\frac{3}{4\pi}} \mathcal{S}_{1m}(\mathbf r_{R'}-\mathbf r_R ) Q_{00}^R  + Q_{1m}^R \; . \label{eq:molecular_dipole_from_atomic_multipoles_with_Slm_dipole}
 \end{align}
\end{subequations}

As the electrostatic multipoles are functions of $\mathbf K$ and $\mathbf r_R$, we can also obtain these quantities for a fragment of a system.
All we have to do is to associate the fragment with a 
``fragment kernel'' $\mathbf K^\mathfrak F\equiv \mathbf{K}\mathbf{S}\mathbf{R}^\mathfrak{F}$, by following the considerations of Sec.~\ref{sec:identification}.
The above definitions must therefore be generalized.
Again restricting ourselves to the case of atomic fragments, this leads to the following definition of the atomic multipoles:
\begin{equation} \label{fragmult}
 Q^A_{\ell m} \equiv \Tr(\mathbf{K}\mathbf{S}\mathbf{R}^A\mathbf{P}^A_{\ell m}) \;.
\end{equation}

With this definition we can now also briefly revisit the projector matrices introduced in Sec.~\ref{sec: Fragment Population Analysis}.
As the monopole matrix is given by $\mathbf{P}_{00}=\mathbf{S}$,
the monopole term for the Mulliken approach 
(Eq.~\eqref{mulli})
reads $Q^A_{00} = \Tr(\mathbf{K}\mathbf{S}\mathbf{T}^A)$,
i.e.\ the trace of $\mathbf{K}\mathbf{S}$ evaluated only for those elements belonging to atom $A$,
 which is indeed nothing other than the well-known Mulliken charge population analysis.
For the L\"owdin approach 
(Eq.~\eqref{loew})
we obtain $Q^A_{00} = \Tr(\mathbf S^{1/2} \mathbf{K}\mathbf{S}^{1/2}\mathbf{T}^A)$,
which indeed corresponds to the L\"owdin charge population analysis.
As $\sum_A \mathbf T^A = \mathbb 1$, both the definitions satisfy the property of Eq.~\eqref{partition},
which is important to ensure the preservation of the total monopole of the system. In other terms, we always have $\sum_A Q_{00}^A = \Tr\left( \mathbf K \mathbf S\right)$.
In the NPA approach the self-duality of the NAO gives $Q^A_{00}=\Tr\left(\mathbf N^A\right)$.

If the fragment is not a single atom, but rather an ensemble of atoms, the projector $\hat W^\mathfrak{F}$ onto that fragment can then simply be defined as the sum over the projectors onto the atoms constituting the fragment, i.e.\ $\hat W^\mathfrak{F}=\sum_{A \in \mathfrak{F}} \hat W^A$.
By linearity and by employing Eq.~\eqref{shiftedmp}, we can obtain the fragment's multipoles in terms of their atomic counterparts:
\begin{equation}
 Q^{\mathfrak{F}}_{\ell m} =  \sum_{\ell'=0}^\ell \sum_{m'=-\ell'}^{\ell'}  \sum_{A \in \mathfrak{F}} Q^A_{\ell' m'} \mathcal C^{\ell m}_{\ell' m'}(\mathbf r_\mathfrak{F} - \mathbf r_A)  \;.
 \label{eq:fragment_multipoles_from_atomic_multipoles}
\end{equation}

 With a fragment projector defined as a sum of atomic projectors we can provide the \emph{atomic} contribution to the electrostatic description of a given fragment. However, such an ``atoms-in-molecule'' description of the fragment must be taken with great care:
Indeed, even if the fragment $\mathfrak{F}$ is reliable in the sense described by Eqs.~\eqref{partition} and \eqref{idempotency}, 
these conditions are in general \emph{not} fulfilled for the atoms $A \in \mathfrak{F}$.
If this is the case the atomic multipoles $Q^A_{\ell m}$ must \emph{not} be considered as (pseudo-) observables as only the fragment as a whole is a reasonable partition of the system.

The above consideration is very important. A charge population analysis may be meaningful for a \emph{molecule}, but not for the atoms belonging to the molecule;
this is due to the fact that the atoms themselves are \emph{not} separable entities of the molecule.
Our criteria allow us to quantify this separability in the basis set used for the population analysis, thereby giving the possibility of associating (or not) such pseudo-observables with well-identified portions of the QM system.

\section{Relation between fragment definition and basis set}
\label{sec: fragment and basis}
We have presented a \emph{quantitative} criterion to identify a fragment within a large system, and we pointed out that its fulfillment does not only depend on the actual fragmentation choice, but also on the nature of the support function basis.
In other terms, the possibility of ``splitting'' a system into fragments is not only an intrinsic property of the system, but also of the set of support functions used to describe it.

Indeed, we have so far avoided any discussion about the specific localized basis set that we use --- rather we simply assumed that a suitable choice exists. In principle there is no constraint on the exact form of the support functions --- 
they can either be a contraction of an underlying basis set, which is the case for example in \textsc{BigDFT}~\cite{mohr-daubechies-2014,mohr-accurate-2015} or 
\textsc{ONETEP}~\cite{skylaris-introducing-2005,haynes-onetep-2006,mostofi-onetep-2007,skylaris-recent-2008},
or predefined atomic basis functions, either numerical or analytic,  as for instance in \textsc{Conquest}~\cite{bowler-practical-2000,bowler-recent-2006,bowler-an-overview-2010},
\textsc{Quickstep}~\cite{vandevondele-quickstep-2005} or \textsc{SIESTA}~\cite{soler-the_siesta-2002,artacho-the_siesta-2008};
an overview over popular electronic structure codes and the basis sets they use can be found in Ref.~\citenum{ratcliff-2016-challenges}.

In this section we want to discuss this important relation between fragment definition and basis set in more detail.
More precisely, we define \textit{a priori} the fragments and the projector matrix, and we discuss the impact of different basis sets in fragments identifications within this setup.
As an illustrative test we take a system where the fragments can readily be identified by chemical intuition, 
namely a droplet of 100 water molecules, 
extracted manually from a larger bulk liquid water system; as it will only serve as a playground, no particular thermalization/relaxation was performed.

\begin{figure*}
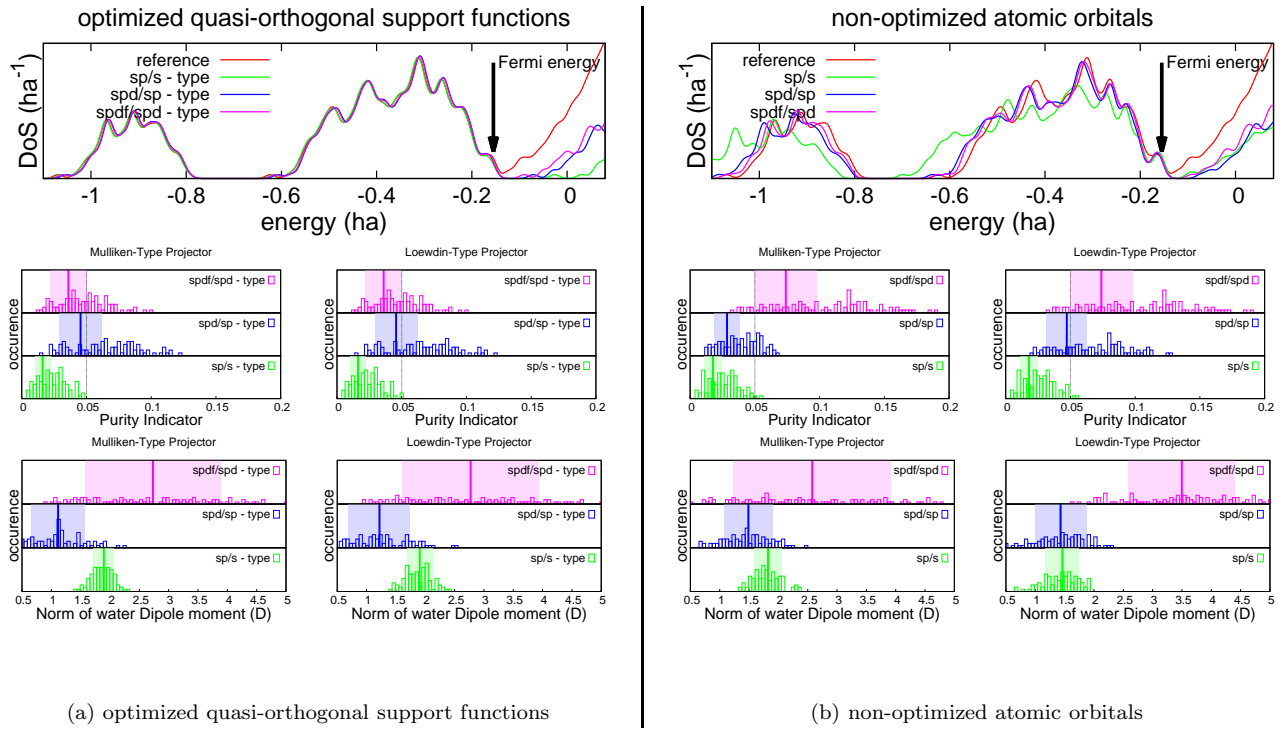

   \subfloat[][optimized quasi-orthogonal support functions]{
   \begin{tabular}{lr}
   \multicolumn{2}{c}{\dossf} \\
  \pisfm & \pisfl \\
  \vspace{0.8cm}
  \dsfm & \dsfl
 \end{tabular}
 \label{fig:waterdroplet_standard}}
   \hspace{.01\textwidth}
   \rulesep
   \hspace{.01\textwidth}
  \subfloat[][non-optimized atomic orbitals]{
  \begin{tabular}{lr}
   \multicolumn{2}{c}{\dosao} \\
  \piaom & \piaol \\
  \vspace{0.8cm}
  \daom & \daol
 \end{tabular}
   \label{fig:waterdroplet_ao}}
   \caption{Comparison of the density of states (first row), purity indicator (second row), and molecular dipoles (third row), for a non-relaxed water droplet consisting of 100 molecules. For the density of states the curves were shifted such that the Fermi energies always coincide with that of the reference calculation, and a Gaussian smearing with $\sigma=\unit[0.27]{eV}$
was applied. For the purity indicator and the dipole moments we present the result for both the Mulliken and L\"owdin approaches. 
The vertical bar at 0.05 in the second panel indicates the ``level of confidence'', i.e.\ we consider a fragment to be reasonable for values below this threshold.
Fig.~\ref{fig:waterdroplet_standard} shows the outcome for the optimized quasi-orthogonal \textsc{BigDFT} support functions, whereas Fig.~\ref{fig:waterdroplet_ao} shows the situation when the support functions are replaced by (unoptimized) atomic orbitals. As can be seen the first case is rather insensitive to the choice of the approach (Mulliken or L\"owdin), whereas the non-orthogonal atomic orbitals show strong deviations. Moreover and most important for this study, the only setup which yields a good result for all measurements is that using the \textit{minimal} set of optimized support functions.}
   \label{fig:waterdroplet}
 \end{figure*}

\paragraph{Basis Set Setups: Optimized Molecular Orbitals \emph{vs.}\ Atomic Orbitals}
In Fig.~\ref{fig:waterdroplet} we present three quantities --- density of states (DoS), purity indicator and molecular dipoles --- for different basis sets. We compare a setup where we use the optimized quasi-orthogonal support functions of \textsc{BigDFT} (Fig.~\ref{fig:waterdroplet_standard}) with a setup where we use non-optimized atomic orbitals (AO, Fig.~\ref{fig:waterdroplet_ao}).
The optimized support functions are obtained by minimizing --- within the underlying Daubechies wavelet~\cite{daubechies-ten-1992} basis of \textsc{BigDFT} --- a target function that ensures both accuracy and locality, and it has been demonstrated that they are capable of representing the KS orbitals and derived quantities well~\cite{mohr-daubechies-2014,mohr-accurate-2015}.
The AO, on the other hand, are obtained by solving --- again within the wavelet basis --- the KS equation for the isolated atom using HGH pseudopotentials~\cite{hartwigsen-relativistic-1998} including a nonlinear core correction~\cite{willand-norm-conserving-2013}.
Both the AO and the optimized basis were confined in localization regions centered around the atoms with a radius of \unit[3.7]{\r{A}} for \ce{H} and \unit[4.0]{\r{A}} for \ce{O},
and the PBE functional~\cite{perdew-generalized-1996} was used. 
\begin{table*}\tablesize
 \begin{tabular}{l l rrr r rrr}
  \toprule
   && \multicolumn{3}{c}{sp/s optimized} && \multicolumn{3}{c}{sp/s atomic orbitals} \\
      \cmidrule{3-5} \cmidrule{7-9}
   && \multicolumn{1}{c}{\ce{H2O}} & \multicolumn{1}{c}{O} & \multicolumn{1}{c}{H} && \multicolumn{1}{c}{\ce{H2O}} & \multicolumn{1}{c}{O} & \multicolumn{1}{c}{H} \\
    Mulliken  && $0.02(1)$ & $0.16(1)$ & $0.45(0)$  &&  $0.03(1)$ & $0.16(1)$ & $0.46(1)$  \\
    L\"owdin  && $0.03(1)$ & $0.16(1)$ & $0.45(0)$  &&  $0.03(1)$ & $0.17(1)$ & $0.48(0)$  \\
    quality   && \multicolumn{1}{c}{\good} & \multicolumn{1}{c}{\bad} & \multicolumn{1}{c}{\bad} && \multicolumn{1}{c}{\good} & \multicolumn{1}{c}{\bad} & \multicolumn{1}{c}{\bad} \\
  \bottomrule
 \end{tabular}
 \caption{Purity indicator of the droplet constituents for the sp/s setup, using the definition of Eq.~\eqref{eq:normalized_purity_indicator}.
  The values for the atoms are considerably larger than those for the entire molecules, indicating that the atoms alone should not be considered as independent fragments.
   }
  \label{tab:purity_waterdroplet_atomic_kernels}
\end{table*}
 \begin{table*}\tablesize
 \begin{tabular}{l l rr r rr}
  \toprule
  && \multicolumn{2}{c}{ sp/s optimized} && \multicolumn{2}{c}{sp/s atomic orbitals} \\
     \cmidrule{3-4} \cmidrule{6-7} 
  && \multicolumn{1}{c}{Mulliken} & \multicolumn{1}{c}{L\"owdin} && \multicolumn{1}{c}{Mulliken} & \multicolumn{1}{c}{L\"owdin} \\
 \ce{H2O} dipole (D) &&  $1.89(18)$ & $1.90(22)$ && $1.83(23)$ & $1.46(29)$  \\
         quality && \multicolumn{1}{c}{\good} & \multicolumn{1}{c}{\good} && \multicolumn{1}{c}{\good} & \multicolumn{1}{c}{\bad} \\
  \bottomrule
 \end{tabular}
 \caption{Mean value of the molecular dipole moment of the droplet molecules, for the sp/s setups of Fig.~\ref{fig:waterdroplet}.
 }
  \label{tab:purity_waterdroplet_kernels}
\end{table*}
 \begin{table*}\tablesize
 \begin{tabular}{l l ccc c ccc}
  \toprule
   && \multicolumn{3}{c}{optimized} && \multicolumn{3}{c}{atomic orbitals} \\
   \cmidrule{3-5} \cmidrule{7-9} 
    && \multicolumn{1}{c}{sp/s}    & \multicolumn{1}{c}{spd/sp} & \multicolumn{1}{c}{spdf/spd} && \multicolumn{1}{c}{sp/s}    & \multicolumn{1}{c}{spd/sp} & \multicolumn{1}{c}{spdf/spd} \\
    DoS && \good & \good & \good && \bad & \good & \good \\
    non-purity && \good & \bad & \bad && \good & \bad & \bad \\
    \ce{H2O} dipole && \good & \bad & \bad && \good/\bad & \bad & \bad \\
  \bottomrule
 \end{tabular}
\caption{Summary of the quality of the description provided by the different setups,
highlighting how the quality of the results potentially depends on the basis setup. Overall only the optimized minimal basis is able to provide reliability in all the categories.}
\label{tab:basis_function_summary}
\end{table*}

For each setup we varied the number of support functions per O/H atom, namely (following the nomenclature of atomic orbitals) of type sp/s, spd/sp and spdf/spd. 
Note that in the augmented setups we did \emph{not} alter the localization regions of the basis, we only included more components.
All setups are compared with a reference calculation done using the cubic scaling version of the \textsc{BigDFT} code~\cite{genovese-daubechies-2008}, which does not use any localization constraints.
For the calculation of the purity indicator and the molecular dipoles we show results for both the Mulliken and L\"owdin approaches, in order to also investigate the effect of the particular choice of the projector matrix.

\paragraph{Description of the Electronic Structure} As can be seen from the uppermost panel of Fig.~\ref{fig:waterdroplet}, all calculations with the optimized functions reproduce
the reference DoS.
The atomic orbitals, on the other hand, exhibit serious deviations for the minimal sp/s basis sets; reasonably accurate results can only be obtained for the larger spd/sp and, even better, spdf/spd setups.
In other words, the basis must be larger, compared to the optimized case, to describe the electronic structure precisely
 --- a fact which is well known from codes which use fixed atomic orbitals~\cite{jensen-atomic-2013}.
 
\paragraph{Purity indicator for the \ce{H2O} molecules} Next we investigate the influence of these different basis sets on the fragment definitions, using the purity indicator derived above.
According to the definition in Eq.~\eqref{eq:normalized_purity_indicator}, a value below a ``level of confidence'' of the order of a few percent seems to be low enough to consider the fragment as a subsystem.
We set from now on our criterion to 5\%; in other terms, we consider the subsystem as a fragment if the projection operator modifies the value of the fragment monopole by no more than 5\%.

As can be seen from the values in the second row of Fig.~\ref{fig:waterdroplet}, the setups using a small basis yield almost pure fragment kernels,
whereas those using a larger basis lead to considerable deviations from zero.

\paragraph{Influence on the measure of the molecular dipoles}  Let us now discuss how this translates into the calculation of the pseudo-observables of the fragments.
In the third panel of Fig.~\ref{fig:waterdroplet}, we plot the distribution of the individual water dipoles within the droplet, calculated as described in Sec.~\ref{sec: multipole decomposition}
We have found the well-known result that the multipole values depend \emph{strongly} on the basis set, even though the DoS is correctly reproduced.
More precisely, we see that --- in an apparently counterintuitive way --- the more ``complete'' the basis set is, the less sound the results for these quantities are.
However, taking into account the results from the second panel, these outcomes become understandable:
For those setups yielding large values for the purity indicator we lose the interpretation of molecular dipoles as (pseudo-)observables.
For the non-minimal basis sets, a Mulliken or L\"owdin analysis 
appears therefore unjustified --- in contrast to the minimal setup, where we get sound values of the molecular dipoles within this (unthermalized) toy droplet.

We should however recall that the purity indicator does not reflect the information about the completeness of the basis set, but only the suitability of the fragment identification \emph{within} the basis. Indeed, for the sp/s setup, the purity indicators are equally good for the optimized and AO setups, as shown in Tab.~\ref{tab:purity_waterdroplet_atomic_kernels}. Nevertheless, for the AO sp/s setup we still have --- as pointed out before --- a too crude representation of the electronic structure of the droplet,  as the KS orbitals are badly expressed in this small AO basis.

\paragraph{Unreliability of atomic multipoles} In Tab.~\ref{tab:purity_waterdroplet_atomic_kernels} we also present the purity indicators considering only the individual atoms as fragments. Compared to the molecules, those values are substantially higher,  indicating that atomic multipoles within a water molecule can \emph{not} be considered as physical observables.
Rather it is necessary to consider a water molecule as a single non-splittable unit, and only the multipole values for this unit can be considered as meaningful and allow a physical interpretation.
We will give another demonstration of the unreliability of atomic multipoles in Sec.~\ref{sec:DNA in water (results)}.

\paragraph{Reliability of Mulliken \emph{vs.} Loewdin} Additionally we also want to emphasize that all results for the optimized support functions are almost invariant under the choice between Mulliken and L\"owdin, whereas the numbers obtained from the atomic orbitals change noticeably.
This is a direct consequence of the quasi-orthogonality of the \textsc{BigDFT} support functions, in contrast to the non-orthogonality of the atomic orbitals. Indeed we see in Tab.~\ref{tab:purity_waterdroplet_kernels}, showing the mean molecular dipole moments, that the AO L\"owdin results are considerably worse than the three other setups. This can be explained by the fact that the L\"owdin approach increases the support of the basis whilst orthogonalizing them, thereby losing the correspondence between orbital and atom.
 
\paragraph{Advantages of Optimized and Minimal (Molecular) basis}
In summary, the purity indicators suggest that \emph{only} the minimal basis setup is meaningful within a Mulliken or L\"owdin approach.
We thus see a clear advantage of using a basis set which is optimized in situ, as indicated by the summary in Tab.~\ref{tab:basis_function_summary}.
Such an optimized minimal basis is \emph{complete enough} for an accurate description of the electronic structure,  but also \emph{small enough} for an accurate description of atomic charges and molecular dipoles.

This advantage of a smaller basis for the characterization of the atomic charges and dipoles might appear counterintuitive. However we have to recall that the richer the basis is the more Rydberg states it contains, making the fragment kernel less pure since both Mulliken and L\"owdin treat all basis functions on an equal footing.
An approach aiming at coping as well with such larger basis sets should thus be able to filter out those basis functions which mainly contribute to the representation of virtual states.
Since neither Mulliken nor L\"owdin have this ability, this implies that these approaches work best -- if not exclusively -- for a minimal basis,
which in turn means that it is indispensable to use an optimized basis set in order to reach a high precision.
The other way around, we see that the use of a minimal and optimized basis allows the usage of simple projectors like Mulliken and L\"owdin, without the need to resort to more involved approaches.

\section{Application to a complex heterogeneous system --- solvated DNA}
\label{sec:DNA in water (results)}
\begin{figure}
 \includegraphics[width=1.0\columnwidth]{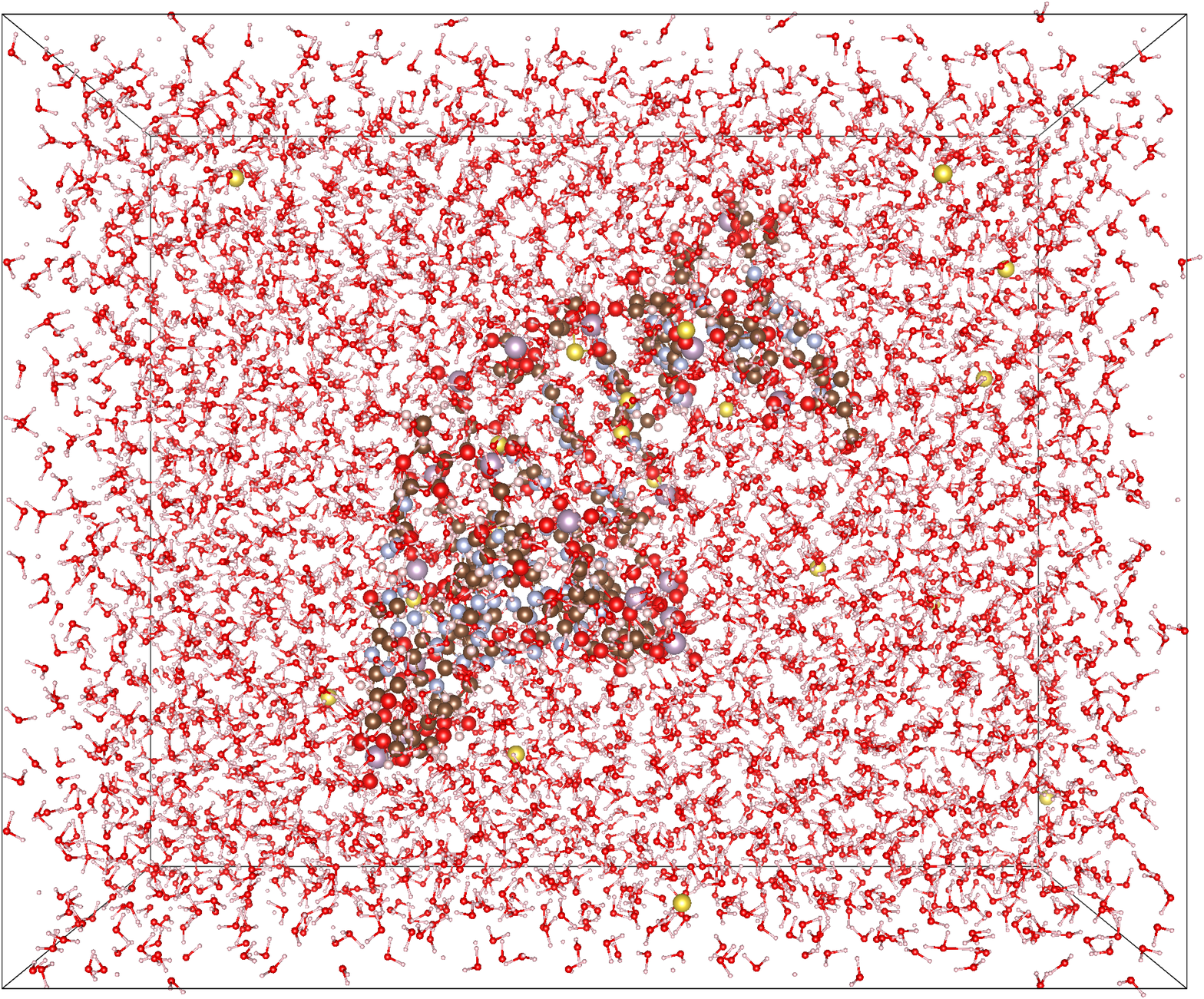}
 \caption{Visualization~\cite{momma-VESTA-2011} of the used DNA fragment (11 base pairs) in Na-water solution, consisting in total of 15,613 atoms.}
 \label{fig:DNA_water-Na}
\end{figure}

In Sec.~\ref{sec: fragment and basis} we have seen that the use of a minimal and optimized basis allows the use of simple population schemes like Mulliken or L\"owdin while still yielding a precise description of the electronic structure.
We now want to apply the developed concepts to a complex heterogeneous system 
where the fragments are not immediately identifiable.
We present results for a rather large system, namely an 11 base pair DNA fragment (made only of Guanine and Cytosine nucleotides)
which is embedded into a sodium-water solution, giving in total 15,613 atoms.
To get a realistic setup we took one snapshot from an extended MD
simulation,
run with Amber 11~\cite{case-the-2005,Amber11} and the ff99SB force field~\cite{hornak-comparison-2006};
the system is shown in Fig.~\ref{fig:DNA_water-Na}.

\begin{figure}
 \subfloat[][Net charges for the individual atoms.]{
  \includegraphics[width=1.0\columnwidth]{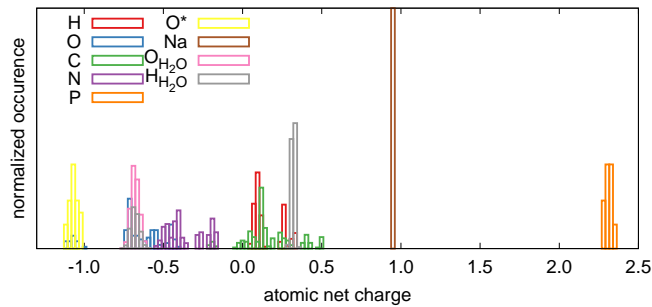}
 \label{fig:charges_fullQM}}
 \\
 \subfloat[][Net charges for some reasonably selected fragments.]{
  \includegraphics[width=1.0\columnwidth]{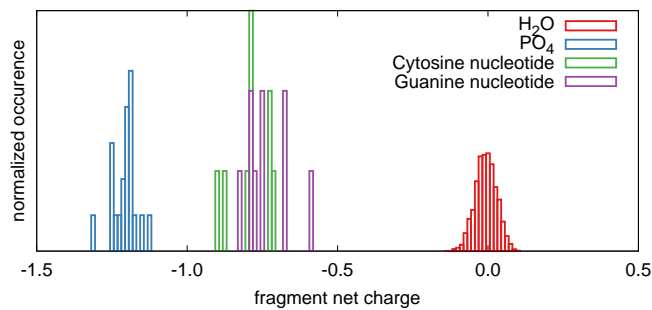}
 \label{fig:charges-fragments_fullQM}}
 \caption{Fragment net charges for the system shown in Fig.~\ref{fig:DNA_water-Na} for various fragment definitions. In Fig.~\ref{fig:charges_fullQM} we chose as fragments the individual atoms, whereas in Fig.~\ref{fig:charges-fragments_fullQM} we chose as fragments groups of atoms, following chemical intuition.}
 \label{fig:charges_DNA}
\end{figure}

In spite of the large dimensions, the linear scaling approach of \textsc{BigDFT}~\cite{mohr-daubechies-2014,mohr-accurate-2015} can easily perform a full QM calculation of the entire system.
Following the considerations of Sec.~\ref{sec: fragment and basis} a minimal set of basis functions was employed.
As a first step we took as candidates for the fragments just the individual atoms; in Fig.~\ref{fig:charges_fullQM} we show the atomic charges that we get from such a fragment definition using the Mulliken projector.

\begin{table*}\tablesize
 \subfloat[][Non-purity for the individual atoms.]{
  \begin{tabular}{l l rrrrrrr}
  \toprule
     && \multicolumn{1}{c}{\ce{H}} & \multicolumn{1}{c}{\ce{C}} & \multicolumn{1}{c}{\ce{N}} & \multicolumn{1}{c}{\ce{O}} & \multicolumn{1}{c}{\ce{Os}} & \multicolumn{1}{c}{\ce{Na}} & \multicolumn{1}{c}{\ce{P}} \\
     \cmidrule{3-9}
   purity indicator && $0.48$ & $0.48$ & $0.32$ & $0.15$ & $0.12$ & $0.04$ & $0.34$ \\
   quality      && \multicolumn{1}{c}{\bad}   & \multicolumn{1}{c}{\bad}   & \multicolumn{1}{c}{\bad}   & \multicolumn{1}{c}{\bad}   & \multicolumn{1}{c}{\bad}   & \multicolumn{1}{c}{\good}  & \multicolumn{1}{c}{\bad}   \\
  \bottomrule
  \end{tabular}
  \label{tab:purity_DNA-atoms}
  }
  \\
 \subfloat[][Non-purity for some reasonably chosen fragments.]{
  \begin{tabular}{l l  rrrr}
  \toprule
     && \multicolumn{1}{c}{\ce{PO4}} & \multicolumn{1}{c}{Cyt} & \multicolumn{1}{c}{Gua} & \multicolumn{1}{c}{\ce{H2O}} \\
     \cmidrule{3-6}
   purity indicator && $0.05$ & $0.01$ & $0.01$ & $0.01$ \\
   quality      && \multicolumn{1}{c}{\good}  & \multicolumn{1}{c}{\good}  & \multicolumn{1}{c}{\good}  & \multicolumn{1}{c}{\good} 	 \\
  \bottomrule
  \label{tab:purity_DNA-fragments}
  \end{tabular}
  }
  \caption{Purity indicator according to Eq.~\eqref{eq:normalized_purity_indicator} (using the Mulliken projector matrix of Eq.~\eqref{mulli}), where the fragment is either a single atom (Tab.~\ref{tab:purity_DNA-atoms}) or --- following chemical intuition --- composed of several atoms (Tab.~\ref{tab:purity_DNA-fragments}).}
\end{table*}

\paragraph{Identification of systems' moieties} To verify whether this fragment choice was sensible, we show in Tab.~\ref{tab:purity_DNA-atoms} the purity indicator for each atom type. As can be seen there are considerable differences, ranging from 4\% for \ce{Na} to 48\% for \ce{H} and \ce{C}.
Once again, this means that for such population methods in this basis care should be taken when extracting atomic charges and multipoles, as in general the atoms cannot 
be considered as ``independent''. As specific examples we focus on the two species which have a large positive net charge, namely \ce{Na} and \ce{P}. The purity indicator for \ce{Na} is very small and thus confirms that \emph{the basis functions employed}
are in line with the  chemical intuition that these \ce{Na} atoms can be considered as ``independent fragments'', 
assuming a fragment selection provided by Mulliken-like projectors.
This also agrees with their chemically sound atomic net charge, which is close to 1.
The purity indicator for \ce{P}, on the other hand, is considerably larger, 
together with a surprisingly high value for its net charge.
This indicates that \ce{P} alone is not an optimal definition of a  fragment in this case.
Indeed the phosphorus atoms are part of a phosphate group \ce{PO4}, 
and if we take this unit as a fragment definition the purity indicator decreases considerably and is 
close to that of \ce{Na}, as shown in Tab.~\ref{tab:purity_DNA-fragments}.
The same scenario also applies for the other atoms of the system. If we consider each full nucleotide within the DNA as a fragment, 
we can see that the purity indicator decreases even further.
The same order of magnitude can be observed for the water molecules, which --- not surprisingly --- again form a reliable fragment.
In Fig.~\ref{fig:charges-fragments_fullQM} we show the fragment charges for these more reasonable fragment choices.
In summary, the purity indicator allows us --- for a given choice of the basis and projector --- to select fragments in an unbiased and reliable way.

\paragraph{Charge population analysis of the DNA nucleotides} The above charge analysis also allows us to determine how much of the \ce{Na} charge has gone to the DNA. 
The 20 \ce{Na} atoms have lost 19.2 electrons (corresponding to an average ionization of 0.96), 
out of which 3.6 have gone to the water and the other 15.6 to the DNA. Considering the aforementioned purity indicators, the 
charge transfer appears to be chemically reliable as it corresponds to a transfer between well-defined fragments.

\section{Conclusions and outlook}
The scope of this paper was to discuss the identification and representation of fragments within large quantum systems.
In particular we aimed to answer the question of under which circumstances the properties of such fragments can be considered as meaningful (pseudo-)observables.
As a basic criterion for the suitability of a fragment definition we identified the purity of the density matrix belonging to the fragment.
This so-called purity indicator is a functional of the fragment projector chosen, the basis set employed, and the fragment considered.

If the value of the  purity indicator is small,
there is only a little coupling between the density operators of the fragment and the system, and the fragment can be considered as a meaningful sub-unit.
In this case it is likely that the characteristics of the fragments can be considered as meaningful observables with a physical interpretation. 
Moreover, the reverse conclusion is even more important: Since a low value of the purity indicator is a \emph{necessary} condition, it will be very difficult to describe the electronic structure of a fragment with meaningful observables --- like for example electrostatic multipoles or partial DoS --- if it does not fulfill this requirement within the given computational setup.

In addition we demonstrated that the use of an \textit{optimized and minimal} localized basis set is of great advantage.
This allows, on the one hand, to correctly identify the fragment 
even for simple projection methods like Mulliken and Loewdin, and on the other hand to describe the electronic structure with a high precision.
Using a larger basis leads to considerably less pure fragment kernels, even for chemically sound fragments such as water molecules within a droplet, and thus renders the entire fragmentation procedure questionable;
using a non-optimized basis requires the use of a large set of functions in order to correctly describe the electronic structure, which in turn leads to the aforementioned fragmentation issues and the need to use more delicate and involved fragment projection methods.
Only the combination of a minimal and optimized basis thus provides satisfying results with respect to both aspects.

Concerning the observables, we focused in this paper on the multipoles of the fragments.
Our formalism 
allows the calculation of multipoles of any order, which is important to provide an accurate description of the fragment's electrostatic potential~\cite{stone-distributed-1981}, in line with established results
based on atomic descriptions~\cite{sokalski-cumulative-1983,williams-representation-1988,sokalski-cumulative-1992,whitehead-transferable-2003,day-beyond-2005,plattner-higher-2009,kramer-deriving-2013}.
They might thus be used in the context of an electrostatic embedding, thereby reducing the complexity of a QM calculation and paving the way towards powerful multiscale calculations~\cite{ratcliff-2016-challenges}.
The use of such electrostatic observables in the context of embedded QM calculations will be considered in a forthcoming publication~\cite{mohr-fragments2-2017}.

\section{Acknowledgments}
We would like to thank Thierry Deutsch for valuable discussions and F\'atima Lucas for providing various test systems and helpful discussions. 
This research used resources of the Argonne Leadership Computing Facility, which is a DOE Office of Science User Facility supported under Contract DE-AC02-06CH11357.
SM acknowledges support from the European Centre of Excellence MaX (project ID 676598).
LG acknowledges support from the EU ExtMOS project (project ID 646176) and the European Centre of Excellence EoCoE (project ID 676629).

\appendix

\section{Definition of the spherical harmonics and translation relations}
\label{sec: Definition of the spherical harmonics and translation relations}
The (real) solid spherical harmonics 
are defined in terms of the corresponding complex functions as (using $\mathbf r \equiv \left( r ,\Omega\right)$):
\begin{multline}
 \mathcal{S}_{\ell m}(r,\Omega) \equiv \\
 \begin{cases} 
\frac{1}{\sqrt 2} r^\ell \left((-1)^m Y_{\ell m}(\Omega) +  Y_{\ell,- m}(\Omega)\right) & m > 0\;,\\
 r^\ell Y_{\ell 0}(\Omega) & m=0\;,\\
\frac{1}{\sqrt 2\mathrm i} r^\ell \left((-1)^mY_{\ell |m|}(\Omega) - Y_{\ell,- |m|}(\Omega)\right) & m < 0\;.
 \end{cases}
\end{multline}
With these conventions they satisfy the orthogonality relation
\begin{equation}
 \int \frac{S_{\ell m}(r,\Omega) S_{\ell' m'}(r,\Omega)}{r^{2\ell}} \, \mathrm d \Omega = \delta_{\ell \ell'}\delta_{m m'} \;,
\end{equation}
for any radial value $r>0$.
The real spherical harmonics satisfy the relation~\cite{RSHshift}
\begin{multline}\label{stranslated}
\mathcal{S}_{\ell m}(\mathbf r+ \mathbf \Delta) = \sum_{\ell'=0}^\ell \sum_{m'=-\ell'}^{\ell'} \mathcal{S}_{\ell' m'}(\mathbf r) \\ \times 
\sqrt{\frac{4\pi}{2\ell'+1}}
\mathcal C^{\ell m}_{\ell' m'}(\mathbf \Delta)\;,
\end{multline}
where the functions
\begin{equation}
\mathcal C^{\ell m}_{\ell' m'}(\mathbf r) = \sqrt{\frac{2\ell'+1}{4\pi}}  \sum_{m''=\ell'-\ell}^{\ell-\ell'}\mathcal{S}_{\ell-\ell' m''}(\mathbf r)  C^{\ell m}_{\ell' m' m''}
\end{equation}
are described in terms of the coeffcients $C^{\ell m}_{\ell' m' m''}$ given by (see also Supplementary Information of Ref.~\citenum{RSHshift}):
\begin{equation}
 \begin{aligned}
  & C^{\ell,m}_{0,0,m''}=\sqrt{4\pi} \delta_{m m''} \;, \\
  & C^{\ell,m}_{\ell,m',m''}= \sqrt{4\pi} \delta_{m' m} \delta_{0 m''} \;, \\
  & C^{2,-2}_{1 m' m''}=\sqrt{4\pi} \sqrt{\frac{5}{3}} \left(\delta_{m',-1} \delta_{m'',1}+\delta_{m',1} \delta_{m'',-1}\right) \;, \\
  & C^{2,-1}_{1 m' m''}=\sqrt{4\pi} \sqrt{\frac{5}{3}} \left(\delta_{m',-1} \delta_{m'',0}+\delta_{m',0} \delta_{m'',-1}\right) \;, \\
  & C^{2,0}_{1 m' m''}=\sqrt{4\pi} \frac{\sqrt{5}}{3} \left(-\delta_{m',-1} \delta_{m'',-1}\right. \\ & \qquad \qquad \quad + \left. 2 \delta_{m',0} \delta_{m'',0} -\delta_{m',1} \delta_{m'',1}\right) \;, \\
  & C^{2,1}_{1 m' m''}=\sqrt{4\pi} \sqrt{\frac{5}{3}} \left(\delta_{m',0} \delta_{m'',1}+\delta_{m',1} \delta_{m'',0}\right) \;, \\
  & C^{2,2}_{1 m' m''}=\sqrt{4\pi} \sqrt{\frac{5}{3}} \left(-\delta_{m',-1} \delta_{m'',-1}+\delta_{m',1} \delta_{m'',1}\right) \;.
 \end{aligned}
 \label{eq:coeffs_for_spherical_harmonics_translation}
\end{equation}

\bibliography{citationlist}

\begin{thebibliography}{59}%
\makeatletter
\providecommand \@ifxundefined [1]{%
 \@ifx{#1\undefined}
}%
\providecommand \@ifnum [1]{%
 \ifnum #1\expandafter \@firstoftwo
 \else \expandafter \@secondoftwo
 \fi
}%
\providecommand \@ifx [1]{%
 \ifx #1\expandafter \@firstoftwo
 \else \expandafter \@secondoftwo
 \fi
}%
\providecommand \natexlab [1]{#1}%
\providecommand \enquote  [1]{``#1''}%
\providecommand \bibnamefont  [1]{#1}%
\providecommand \bibfnamefont [1]{#1}%
\providecommand \citenamefont [1]{#1}%
\providecommand \href@noop [0]{\@secondoftwo}%
\providecommand \href [0]{\begingroup \@sanitize@url \@href}%
\providecommand \@href[1]{\@@startlink{#1}\@@href}%
\providecommand \@@href[1]{\endgroup#1\@@endlink}%
\providecommand \@sanitize@url [0]{\catcode `\\12\catcode `\$12\catcode
  `\&12\catcode `\#12\catcode `\^12\catcode `\_12\catcode `\%12\relax}%
\providecommand \@@startlink[1]{}%
\providecommand \@@endlink[0]{}%
\providecommand \url  [0]{\begingroup\@sanitize@url \@url }%
\providecommand \@url [1]{\endgroup\@href {#1}{\urlprefix }}%
\providecommand \urlprefix  [0]{URL }%
\providecommand \Eprint [0]{\href }%
\providecommand \doibase [0]{http://dx.doi.org/}%
\providecommand \selectlanguage [0]{\@gobble}%
\providecommand \bibinfo  [0]{\@secondoftwo}%
\providecommand \bibfield  [0]{\@secondoftwo}%
\providecommand \translation [1]{[#1]}%
\providecommand \BibitemOpen [0]{}%
\providecommand \bibitemStop [0]{}%
\providecommand \bibitemNoStop [0]{.\EOS\space}%
\providecommand \EOS [0]{\spacefactor3000\relax}%
\providecommand \BibitemShut  [1]{\csname bibitem#1\endcsname}%
\let\auto@bib@innerbib\@empty
\bibitem [{\citenamefont {Hohenberg}\ and\ \citenamefont
  {Kohn}(1964)}]{hohenberg-inhomogeneous-1964}%
  \BibitemOpen
  \bibfield  {author} {\bibinfo {author} {\bibfnamefont {P.}~\bibnamefont
  {Hohenberg}}\ and\ \bibinfo {author} {\bibfnamefont {W.}~\bibnamefont
  {Kohn}},\ }\bibfield  {title} {\enquote {\bibinfo {title} {Inhomogeneous
  electron gas},}\ }\href {\doibase 10.1103/PhysRev.136.B864} {\bibfield
  {journal} {\bibinfo  {journal} {Phys. Rev.}\ }\textbf {\bibinfo {volume}
  {136}},\ \bibinfo {pages} {B864} (\bibinfo {year} {1964})}\BibitemShut
  {NoStop}%
\bibitem [{\citenamefont {Kohn}\ and\ \citenamefont
  {Sham}(1965)}]{kohn-self_consistent-1965}%
  \BibitemOpen
  \bibfield  {author} {\bibinfo {author} {\bibfnamefont {W.}~\bibnamefont
  {Kohn}}\ and\ \bibinfo {author} {\bibfnamefont {L.~J.}\ \bibnamefont
  {Sham}},\ }\bibfield  {title} {\enquote {\bibinfo {title} {Self-consistent
  equations including exchange and correlation effects},}\ }\href {\doibase
  10.1103/PhysRev.140.A1133} {\bibfield  {journal} {\bibinfo  {journal} {Phys.
  Rev.}\ }\textbf {\bibinfo {volume} {140}},\ \bibinfo {pages} {A1133}
  (\bibinfo {year} {1965})}\BibitemShut {NoStop}%
\bibitem [{\citenamefont {Ratcliff}\ \emph {et~al.}(2017)\citenamefont
  {Ratcliff}, \citenamefont {Mohr}, \citenamefont {Huhs}, \citenamefont
  {Deutsch}, \citenamefont {Masella},\ and\ \citenamefont
  {Genovese}}]{ratcliff-2016-challenges}%
  \BibitemOpen
  \bibfield  {author} {\bibinfo {author} {\bibfnamefont {Laura~E.}\
  \bibnamefont {Ratcliff}}, \bibinfo {author} {\bibfnamefont {Stephan}\
  \bibnamefont {Mohr}}, \bibinfo {author} {\bibfnamefont {Georg}\ \bibnamefont
  {Huhs}}, \bibinfo {author} {\bibfnamefont {Thierry}\ \bibnamefont {Deutsch}},
  \bibinfo {author} {\bibfnamefont {Michel}\ \bibnamefont {Masella}}, \ and\
  \bibinfo {author} {\bibfnamefont {Luigi}\ \bibnamefont {Genovese}},\
  }\bibfield  {title} {\enquote {\bibinfo {title} {Challenges in large scale
  quantum mechanical calculations},}\ }\href {\doibase 10.1002/wcms.1290}
  {\bibfield  {journal} {\bibinfo  {journal} {Wiley Interdiscip. Rev.-Comput.
  Mol. Sci.}\ }\textbf {\bibinfo {volume} {7}},\ \bibinfo {pages} {e1290}
  (\bibinfo {year} {2017})},\ \bibinfo {note} {e1290}\BibitemShut {NoStop}%
\bibitem [{\citenamefont {Gordon}\ \emph {et~al.}(2012)\citenamefont {Gordon},
  \citenamefont {Fedorov}, \citenamefont {Pruitt},\ and\ \citenamefont
  {Slipchenko}}]{gordon-fragmentation-2012}%
  \BibitemOpen
  \bibfield  {author} {\bibinfo {author} {\bibfnamefont {Mark~S.}\ \bibnamefont
  {Gordon}}, \bibinfo {author} {\bibfnamefont {Dmitri~G.}\ \bibnamefont
  {Fedorov}}, \bibinfo {author} {\bibfnamefont {Spencer~R.}\ \bibnamefont
  {Pruitt}}, \ and\ \bibinfo {author} {\bibfnamefont {Lyudmila~V.}\
  \bibnamefont {Slipchenko}},\ }\bibfield  {title} {\enquote {\bibinfo {title}
  {Fragmentation methods: A route to accurate calculations on large systems},}\
  }\href {\doibase 10.1021/cr200093j} {\bibfield  {journal} {\bibinfo
  {journal} {Chem. Rev.}\ }\textbf {\bibinfo {volume} {112}},\ \bibinfo {pages}
  {632} (\bibinfo {year} {2012})},\ \bibinfo {note} {pMID: 21866983},\ \Eprint
  {http://arxiv.org/abs/http://dx.doi.org/10.1021/cr200093j}
  {http://dx.doi.org/10.1021/cr200093j} \BibitemShut {NoStop}%
\bibitem [{\citenamefont {Collins}\ and\ \citenamefont
  {Bettens}(2015)}]{collins-energy-based-2015}%
  \BibitemOpen
  \bibfield  {author} {\bibinfo {author} {\bibfnamefont {Michael~A.}\
  \bibnamefont {Collins}}\ and\ \bibinfo {author} {\bibfnamefont {Ryan P.~A.}\
  \bibnamefont {Bettens}},\ }\bibfield  {title} {\enquote {\bibinfo {title}
  {Energy-based molecular fragmentation methods},}\ }\href {\doibase
  10.1021/cr500455b} {\bibfield  {journal} {\bibinfo  {journal} {Chem. Rev.}\
  }\textbf {\bibinfo {volume} {115}},\ \bibinfo {pages} {5607} (\bibinfo {year}
  {2015})},\ \bibinfo {note} {pMID: 25843427},\ \Eprint
  {http://arxiv.org/abs/http://dx.doi.org/10.1021/cr500455b}
  {http://dx.doi.org/10.1021/cr500455b} \BibitemShut {NoStop}%
\bibitem [{\citenamefont {Wiberg}\ and\ \citenamefont
  {Rablen}(1993)}]{wiberg-comparison-1993}%
  \BibitemOpen
  \bibfield  {author} {\bibinfo {author} {\bibfnamefont {Kenneth~B.}\
  \bibnamefont {Wiberg}}\ and\ \bibinfo {author} {\bibfnamefont {Paul~R.}\
  \bibnamefont {Rablen}},\ }\bibfield  {title} {\enquote {\bibinfo {title}
  {Comparison of atomic charges derived via different procedures},}\ }\href
  {\doibase 10.1002/jcc.540141213} {\bibfield  {journal} {\bibinfo  {journal}
  {J. Comput. Chem.}\ }\textbf {\bibinfo {volume} {14}},\ \bibinfo {pages}
  {1504} (\bibinfo {year} {1993})}\BibitemShut {NoStop}%
\bibitem [{\citenamefont {Fonseca~Guerra}\ \emph {et~al.}(2004)\citenamefont
  {Fonseca~Guerra}, \citenamefont {Handgraaf}, \citenamefont {Baerends},\ and\
  \citenamefont {Bickelhaupt}}]{fonseca-voronoi-2004}%
  \BibitemOpen
  \bibfield  {author} {\bibinfo {author} {\bibfnamefont {C\'elia}\ \bibnamefont
  {Fonseca~Guerra}}, \bibinfo {author} {\bibfnamefont {Jan-Willem}\
  \bibnamefont {Handgraaf}}, \bibinfo {author} {\bibfnamefont {Evert~Jan}\
  \bibnamefont {Baerends}}, \ and\ \bibinfo {author} {\bibfnamefont
  {F.~Matthias}\ \bibnamefont {Bickelhaupt}},\ }\bibfield  {title} {\enquote
  {\bibinfo {title} {Voronoi deformation density (vdd) charges: Assessment of
  the mulliken, bader, hirshfeld, weinhold, and vdd methods for charge
  analysis},}\ }\href {\doibase 10.1002/jcc.10351} {\bibfield  {journal}
  {\bibinfo  {journal} {J. Comput. Chem.}\ }\textbf {\bibinfo {volume} {25}},\
  \bibinfo {pages} {189} (\bibinfo {year} {2004})}\BibitemShut {NoStop}%
\bibitem [{\citenamefont {Choi}\ and\ \citenamefont
  {Fedorov}(2012)}]{choi-reducing-2012}%
  \BibitemOpen
  \bibfield  {author} {\bibinfo {author} {\bibfnamefont {Cheol~Ho}\
  \bibnamefont {Choi}}\ and\ \bibinfo {author} {\bibfnamefont {Dmitri~G.}\
  \bibnamefont {Fedorov}},\ }\bibfield  {title} {\enquote {\bibinfo {title}
  {Reducing the scaling of the fragment molecular orbital method using the
  multipole method},}\ }\href {\doibase
  http://dx.doi.org/10.1016/j.cplett.2012.06.018} {\bibfield  {journal}
  {\bibinfo  {journal} {Chem. Phys. Lett.}\ }\textbf {\bibinfo {volume}
  {543}},\ \bibinfo {pages} {159} (\bibinfo {year} {2012})}\BibitemShut
  {NoStop}%
\bibitem [{\citenamefont {Kitaura}\ \emph {et~al.}(1999)\citenamefont
  {Kitaura}, \citenamefont {Ikeo}, \citenamefont {Asada}, \citenamefont
  {Nakano},\ and\ \citenamefont {Uebayasi}}]{kitaura-fragment-1999}%
  \BibitemOpen
  \bibfield  {author} {\bibinfo {author} {\bibfnamefont {Kazuo}\ \bibnamefont
  {Kitaura}}, \bibinfo {author} {\bibfnamefont {Eiji}\ \bibnamefont {Ikeo}},
  \bibinfo {author} {\bibfnamefont {Toshio}\ \bibnamefont {Asada}}, \bibinfo
  {author} {\bibfnamefont {Tatsuya}\ \bibnamefont {Nakano}}, \ and\ \bibinfo
  {author} {\bibfnamefont {Masami}\ \bibnamefont {Uebayasi}},\ }\bibfield
  {title} {\enquote {\bibinfo {title} {Fragment molecular orbital method: an
  approximate computational method for large molecules},}\ }\href {\doibase
  http://dx.doi.org/10.1016/S0009-2614(99)00874-X} {\bibfield  {journal}
  {\bibinfo  {journal} {Chem. Phys. Lett.}\ }\textbf {\bibinfo {volume}
  {313}},\ \bibinfo {pages} {701} (\bibinfo {year} {1999})}\BibitemShut
  {NoStop}%
\bibitem [{\citenamefont {Fedorov}\ and\ \citenamefont
  {Kitaura}(2007)}]{fedorov-extending-2007}%
  \BibitemOpen
  \bibfield  {author} {\bibinfo {author} {\bibfnamefont {Dmitri~G.}\
  \bibnamefont {Fedorov}}\ and\ \bibinfo {author} {\bibfnamefont {Kazuo}\
  \bibnamefont {Kitaura}},\ }\bibfield  {title} {\enquote {\bibinfo {title}
  {Extending the power of quantum chemistry to large systems with the fragment
  molecular orbital method},}\ }\href {\doibase 10.1021/jp0716740} {\bibfield
  {journal} {\bibinfo  {journal} {J. Phys. Chem. A}\ }\textbf {\bibinfo
  {volume} {111}},\ \bibinfo {pages} {6904} (\bibinfo {year} {2007})},\
  \bibinfo {note} {pMID: 17511437},\ \Eprint
  {http://arxiv.org/abs/http://dx.doi.org/10.1021/jp0716740}
  {http://dx.doi.org/10.1021/jp0716740} \BibitemShut {NoStop}%
\bibitem [{\citenamefont {Gao}(1997)}]{gao-toward-1997}%
  \BibitemOpen
  \bibfield  {author} {\bibinfo {author} {\bibfnamefont {Jiali}\ \bibnamefont
  {Gao}},\ }\bibfield  {title} {\enquote {\bibinfo {title} {Toward a molecular
  orbital derived empirical potential for liquid simulations},}\ }\href
  {\doibase 10.1021/jp962833a} {\bibfield  {journal} {\bibinfo  {journal} {J.
  Phys. Chem. B}\ }\textbf {\bibinfo {volume} {101}},\ \bibinfo {pages} {657}
  (\bibinfo {year} {1997})},\ \Eprint
  {http://arxiv.org/abs/http://dx.doi.org/10.1021/jp962833a}
  {http://dx.doi.org/10.1021/jp962833a} \BibitemShut {NoStop}%
\bibitem [{\citenamefont {Gao}(1998)}]{gao-a-molecular-1998}%
  \BibitemOpen
  \bibfield  {author} {\bibinfo {author} {\bibfnamefont {Jiali}\ \bibnamefont
  {Gao}},\ }\bibfield  {title} {\enquote {\bibinfo {title} {A molecular-orbital
  derived polarization potential for liquid water},}\ }\href {\doibase
  10.1063/1.476802} {\bibfield  {journal} {\bibinfo  {journal} {J. Chem.
  Phys.}\ }\textbf {\bibinfo {volume} {109}},\ \bibinfo {pages} {2346}
  (\bibinfo {year} {1998})},\ \Eprint
  {http://arxiv.org/abs/http://dx.doi.org/10.1063/1.476802}
  {http://dx.doi.org/10.1063/1.476802} \BibitemShut {NoStop}%
\bibitem [{\citenamefont {Wierzchowski}\ \emph {et~al.}(2003)\citenamefont
  {Wierzchowski}, \citenamefont {Kofke},\ and\ \citenamefont
  {Gao}}]{wierzchowski-hydrogen-2003}%
  \BibitemOpen
  \bibfield  {author} {\bibinfo {author} {\bibfnamefont {Scott~J.}\
  \bibnamefont {Wierzchowski}}, \bibinfo {author} {\bibfnamefont {David~A.}\
  \bibnamefont {Kofke}}, \ and\ \bibinfo {author} {\bibfnamefont {Jiali}\
  \bibnamefont {Gao}},\ }\bibfield  {title} {\enquote {\bibinfo {title}
  {Hydrogen fluoride phase behavior and molecular structure: A qm/mm potential
  model approach},}\ }\href {\doibase 10.1063/1.1607919} {\bibfield  {journal}
  {\bibinfo  {journal} {J. Chem. Phys.}\ }\textbf {\bibinfo {volume} {119}},\
  \bibinfo {pages} {7365} (\bibinfo {year} {2003})},\ \Eprint
  {http://arxiv.org/abs/http://dx.doi.org/10.1063/1.1607919}
  {http://dx.doi.org/10.1063/1.1607919} \BibitemShut {NoStop}%
\bibitem [{\citenamefont {Xie}\ and\ \citenamefont
  {Gao}(2007)}]{xie-design-2007}%
  \BibitemOpen
  \bibfield  {author} {\bibinfo {author} {\bibfnamefont {Wangshen}\
  \bibnamefont {Xie}}\ and\ \bibinfo {author} {\bibfnamefont {Jiali}\
  \bibnamefont {Gao}},\ }\bibfield  {title} {\enquote {\bibinfo {title} {Design
  of a next generation force field:  the x-pol potential},}\ }\href {\doibase
  10.1021/ct700167b} {\bibfield  {journal} {\bibinfo  {journal} {J. Chem.
  Theory Comput.}\ }\textbf {\bibinfo {volume} {3}},\ \bibinfo {pages} {1890}
  (\bibinfo {year} {2007})},\ \bibinfo {note} {pMID: 18985172},\ \Eprint
  {http://arxiv.org/abs/http://dx.doi.org/10.1021/ct700167b}
  {http://dx.doi.org/10.1021/ct700167b} \BibitemShut {NoStop}%
\bibitem [{\citenamefont {Xie}\ \emph {et~al.}(2008{\natexlab{a}})\citenamefont
  {Xie}, \citenamefont {Song}, \citenamefont {Truhlar},\ and\ \citenamefont
  {Gao}}]{xie-the-variational-2008}%
  \BibitemOpen
  \bibfield  {author} {\bibinfo {author} {\bibfnamefont {Wangshen}\
  \bibnamefont {Xie}}, \bibinfo {author} {\bibfnamefont {Lingchun}\
  \bibnamefont {Song}}, \bibinfo {author} {\bibfnamefont {Donald~G.}\
  \bibnamefont {Truhlar}}, \ and\ \bibinfo {author} {\bibfnamefont {Jiali}\
  \bibnamefont {Gao}},\ }\bibfield  {title} {\enquote {\bibinfo {title} {The
  variational explicit polarization potential and analytical first derivative
  of energy: Towards a next generation force field},}\ }\href {\doibase
  10.1063/1.2936122} {\bibfield  {journal} {\bibinfo  {journal} {J. Chem.
  Phys.}\ }\textbf {\bibinfo {volume} {128}},\ \bibinfo {pages} {234108}
  (\bibinfo {year} {2008}{\natexlab{a}})},\ \Eprint
  {http://arxiv.org/abs/http://dx.doi.org/10.1063/1.2936122}
  {http://dx.doi.org/10.1063/1.2936122} \BibitemShut {NoStop}%
\bibitem [{\citenamefont {Xie}\ \emph {et~al.}(2008{\natexlab{b}})\citenamefont
  {Xie}, \citenamefont {Song}, \citenamefont {Truhlar},\ and\ \citenamefont
  {Gao}}]{xie-incorporation-2008}%
  \BibitemOpen
  \bibfield  {author} {\bibinfo {author} {\bibfnamefont {Wangshen}\
  \bibnamefont {Xie}}, \bibinfo {author} {\bibfnamefont {Lingchun}\
  \bibnamefont {Song}}, \bibinfo {author} {\bibfnamefont {Donald~G.}\
  \bibnamefont {Truhlar}}, \ and\ \bibinfo {author} {\bibfnamefont {Jiali}\
  \bibnamefont {Gao}},\ }\bibfield  {title} {\enquote {\bibinfo {title}
  {Incorporation of a qm/mm buffer zone in the variational double
  self-consistent field method},}\ }\href {\doibase 10.1021/jp804512f}
  {\bibfield  {journal} {\bibinfo  {journal} {J. Phys. Chem. B}\ }\textbf
  {\bibinfo {volume} {112}},\ \bibinfo {pages} {14124} (\bibinfo {year}
  {2008}{\natexlab{b}})},\ \bibinfo {note} {pMID: 18937511},\ \Eprint
  {http://arxiv.org/abs/http://dx.doi.org/10.1021/jp804512f}
  {http://dx.doi.org/10.1021/jp804512f} \BibitemShut {NoStop}%
\bibitem [{\citenamefont {Wu}\ \emph {et~al.}(2011)\citenamefont {Wu},
  \citenamefont {Liu}, \citenamefont {Zhang},\ and\ \citenamefont
  {Li}}]{wu-linear-scaling-2011}%
  \BibitemOpen
  \bibfield  {author} {\bibinfo {author} {\bibfnamefont {Fangqin}\ \bibnamefont
  {Wu}}, \bibinfo {author} {\bibfnamefont {Wenjian}\ \bibnamefont {Liu}},
  \bibinfo {author} {\bibfnamefont {Yong}\ \bibnamefont {Zhang}}, \ and\
  \bibinfo {author} {\bibfnamefont {Zhendong}\ \bibnamefont {Li}},\ }\bibfield
  {title} {\enquote {\bibinfo {title} {Linear-scaling time-dependent density
  functional theory based on the idea of “from fragments to molecule”},}\
  }\href {\doibase 10.1021/ct200225v} {\bibfield  {journal} {\bibinfo
  {journal} {J. Chem. Theory Comput.}\ }\textbf {\bibinfo {volume} {7}},\
  \bibinfo {pages} {3643} (\bibinfo {year} {2011})},\ \bibinfo {note} {pMID:
  26598260},\ \Eprint
  {http://arxiv.org/abs/http://dx.doi.org/10.1021/ct200225v}
  {http://dx.doi.org/10.1021/ct200225v} \BibitemShut {NoStop}%
\bibitem [{\citenamefont {Liu}\ \emph {et~al.}(2014)\citenamefont {Liu},
  \citenamefont {Zhang},\ and\ \citenamefont {Liu}}]{liu-photoexcitation-2014}%
  \BibitemOpen
  \bibfield  {author} {\bibinfo {author} {\bibfnamefont {Junzi}\ \bibnamefont
  {Liu}}, \bibinfo {author} {\bibfnamefont {Yong}\ \bibnamefont {Zhang}}, \
  and\ \bibinfo {author} {\bibfnamefont {Wenjian}\ \bibnamefont {Liu}},\
  }\bibfield  {title} {\enquote {\bibinfo {title} {Photoexcitation of
  light-harvesting c–p–c60 triads: A flmo-td-dft study},}\ }\href {\doibase
  10.1021/ct500066t} {\bibfield  {journal} {\bibinfo  {journal} {J. Chem.
  Theory Comput.}\ }\textbf {\bibinfo {volume} {10}},\ \bibinfo {pages} {2436}
  (\bibinfo {year} {2014})},\ \bibinfo {note} {pMID: 26580764},\ \Eprint
  {http://arxiv.org/abs/http://dx.doi.org/10.1021/ct500066t}
  {http://dx.doi.org/10.1021/ct500066t} \BibitemShut {NoStop}%
\bibitem [{\citenamefont {Li}\ \emph {et~al.}(2014)\citenamefont {Li},
  \citenamefont {Li}, \citenamefont {Suo},\ and\ \citenamefont
  {Liu}}]{li-localization-2014}%
  \BibitemOpen
  \bibfield  {author} {\bibinfo {author} {\bibfnamefont {Zhendong}\
  \bibnamefont {Li}}, \bibinfo {author} {\bibfnamefont {Hongyang}\ \bibnamefont
  {Li}}, \bibinfo {author} {\bibfnamefont {Bingbing}\ \bibnamefont {Suo}}, \
  and\ \bibinfo {author} {\bibfnamefont {Wenjian}\ \bibnamefont {Liu}},\
  }\bibfield  {title} {\enquote {\bibinfo {title} {Localization of molecular
  orbitals: From fragments to molecule},}\ }\href {\doibase 10.1021/ar500082t}
  {\bibfield  {journal} {\bibinfo  {journal} {Acc. Chem. Res.}\ }\textbf
  {\bibinfo {volume} {47}},\ \bibinfo {pages} {2758} (\bibinfo {year}
  {2014})},\ \bibinfo {note} {pMID: 25019464},\ \Eprint
  {http://arxiv.org/abs/http://dx.doi.org/10.1021/ar500082t}
  {http://dx.doi.org/10.1021/ar500082t} \BibitemShut {NoStop}%
\bibitem [{\citenamefont {Li}\ \emph {et~al.}(2017)\citenamefont {Li},
  \citenamefont {Liu},\ and\ \citenamefont {Suo}}]{li-localization-2017}%
  \BibitemOpen
  \bibfield  {author} {\bibinfo {author} {\bibfnamefont {Hongyang}\
  \bibnamefont {Li}}, \bibinfo {author} {\bibfnamefont {Wenjian}\ \bibnamefont
  {Liu}}, \ and\ \bibinfo {author} {\bibfnamefont {Bingbing}\ \bibnamefont
  {Suo}},\ }\bibfield  {title} {\enquote {\bibinfo {title} {Localization of
  open-shell molecular orbitals via least change from fragments to molecule},}\
  }\href {\doibase 10.1063/1.4977929} {\bibfield  {journal} {\bibinfo
  {journal} {J. Chem. Phys.}\ }\textbf {\bibinfo {volume} {146}},\ \bibinfo
  {pages} {104104} (\bibinfo {year} {2017})},\ \Eprint
  {http://arxiv.org/abs/http://dx.doi.org/10.1063/1.4977929}
  {http://dx.doi.org/10.1063/1.4977929} \BibitemShut {NoStop}%
\bibitem [{\citenamefont {Hern\'andez}\ and\ \citenamefont
  {Gillan}(1995)}]{hernandez-self-1995}%
  \BibitemOpen
  \bibfield  {author} {\bibinfo {author} {\bibfnamefont {E.}~\bibnamefont
  {Hern\'andez}}\ and\ \bibinfo {author} {\bibfnamefont {M.~J.}\ \bibnamefont
  {Gillan}},\ }\bibfield  {title} {\enquote {\bibinfo {title} {Self-consistent
  first-principles technique with linear scaling},}\ }\href {\doibase
  10.1103/PhysRevB.51.10157} {\bibfield  {journal} {\bibinfo  {journal} {Phys.
  Rev. B}\ }\textbf {\bibinfo {volume} {51}},\ \bibinfo {pages} {10157}
  (\bibinfo {year} {1995})}\BibitemShut {NoStop}%
\bibitem [{\citenamefont {Skylaris}\ \emph {et~al.}(2005)\citenamefont
  {Skylaris}, \citenamefont {Haynes}, \citenamefont {Mostofi},\ and\
  \citenamefont {Payne}}]{skylaris-introducing-2005}%
  \BibitemOpen
  \bibfield  {author} {\bibinfo {author} {\bibfnamefont {Chris-Kriton}\
  \bibnamefont {Skylaris}}, \bibinfo {author} {\bibfnamefont {Peter~D.}\
  \bibnamefont {Haynes}}, \bibinfo {author} {\bibfnamefont {Arash~A.}\
  \bibnamefont {Mostofi}}, \ and\ \bibinfo {author} {\bibfnamefont {Mike~C.}\
  \bibnamefont {Payne}},\ }\bibfield  {title} {\enquote {\bibinfo {title}
  {Introducing onetep: Linear-scaling density functional simulations on
  parallel computers},}\ }\href@noop {} {\bibfield  {journal} {\bibinfo
  {journal} {J. Chem. Phys.}\ }\textbf {\bibinfo {volume} {122}},\ \bibinfo
  {eid} {084119} (\bibinfo {year} {2005})}\BibitemShut {NoStop}%
\bibitem [{\citenamefont {Bowler}\ \emph {et~al.}(2006)\citenamefont {Bowler},
  \citenamefont {Choudhury}, \citenamefont {Gillan},\ and\ \citenamefont
  {Miyazaki}}]{bowler-recent-2006}%
  \BibitemOpen
  \bibfield  {author} {\bibinfo {author} {\bibfnamefont {D.~R.}\ \bibnamefont
  {Bowler}}, \bibinfo {author} {\bibfnamefont {R.}~\bibnamefont {Choudhury}},
  \bibinfo {author} {\bibfnamefont {M.~J.}\ \bibnamefont {Gillan}}, \ and\
  \bibinfo {author} {\bibfnamefont {T.}~\bibnamefont {Miyazaki}},\ }\bibfield
  {title} {\enquote {\bibinfo {title} {Recent progress with large-scale ab
  initio calculations: the conquest code},}\ }\href {\doibase
  10.1002/pssb.200541386} {\bibfield  {journal} {\bibinfo  {journal} {Phys.
  Status Solidi B}\ }\textbf {\bibinfo {volume} {243}},\ \bibinfo {pages} {989}
  (\bibinfo {year} {2006})}\BibitemShut {NoStop}%
\bibitem [{mul(1955)}]{mulliken-electronic-1955}%
  \BibitemOpen
  \bibfield  {title} {\enquote {\bibinfo {title} {Electronic population
  analysis on lcao-mo molecular wave functions. i},}\ }\href {\doibase
  10.1063/1.1740588} {\bibfield  {journal} {\bibinfo  {journal} {J. Chem.
  Phys.}\ }\textbf {\bibinfo {volume} {23}},\ \bibinfo {pages} {1833} (\bibinfo
  {year} {1955})},\ \Eprint
  {http://arxiv.org/abs/http://dx.doi.org/10.1063/1.1740588}
  {http://dx.doi.org/10.1063/1.1740588} \BibitemShut {NoStop}%
\bibitem [{\citenamefont {Reed}\ \emph {et~al.}(1985)\citenamefont {Reed},
  \citenamefont {Weinstock},\ and\ \citenamefont
  {Weinhold}}]{reed-natural-1985}%
  \BibitemOpen
  \bibfield  {author} {\bibinfo {author} {\bibfnamefont {Alan~E.}\ \bibnamefont
  {Reed}}, \bibinfo {author} {\bibfnamefont {Robert~B.}\ \bibnamefont
  {Weinstock}}, \ and\ \bibinfo {author} {\bibfnamefont {Frank}\ \bibnamefont
  {Weinhold}},\ }\bibfield  {title} {\enquote {\bibinfo {title} {Natural
  population analysis},}\ }\href {\doibase http://dx.doi.org/10.1063/1.449486}
  {\bibfield  {journal} {\bibinfo  {journal} {J. Chem. Phys.}\ }\textbf
  {\bibinfo {volume} {83}},\ \bibinfo {pages} {735} (\bibinfo {year}
  {1985})}\BibitemShut {NoStop}%
\bibitem [{\citenamefont {Politzer}\ and\ \citenamefont
  {Mulliken}(1971)}]{politzer-comparison-1971}%
  \BibitemOpen
  \bibfield  {author} {\bibinfo {author} {\bibfnamefont {Peter}\ \bibnamefont
  {Politzer}}\ and\ \bibinfo {author} {\bibfnamefont {Robert~S.}\ \bibnamefont
  {Mulliken}},\ }\bibfield  {title} {\enquote {\bibinfo {title} {Comparison of
  two atomic charge definitions, as applied to the hydrogen fluoride
  molecule},}\ }\href {\doibase http://dx.doi.org/10.1063/1.1675638} {\bibfield
   {journal} {\bibinfo  {journal} {J. Chem. Phys.}\ }\textbf {\bibinfo {volume}
  {55}},\ \bibinfo {pages} {5135} (\bibinfo {year} {1971})}\BibitemShut
  {NoStop}%
\bibitem [{\citenamefont {L\"owdin}(1950)}]{loewdin-on-1950}%
  \BibitemOpen
  \bibfield  {author} {\bibinfo {author} {\bibfnamefont {Per-Olov}\
  \bibnamefont {L\"owdin}},\ }\bibfield  {title} {\enquote {\bibinfo {title}
  {On the non-orthogonality problem connected with the use of atomic wave
  functions in the theory of molecules and crystals},}\ }\href {\doibase
  http://dx.doi.org/10.1063/1.1747632} {\bibfield  {journal} {\bibinfo
  {journal} {J. Chem. Phys.}\ }\textbf {\bibinfo {volume} {18}},\ \bibinfo
  {pages} {365} (\bibinfo {year} {1950})}\BibitemShut {NoStop}%
\bibitem [{\citenamefont {L\"owdin}(1970)}]{loewdin-on-1970}%
  \BibitemOpen
  \bibfield  {author} {\bibinfo {author} {\bibfnamefont {Per-Olov}\
  \bibnamefont {L\"owdin}},\ }\bibfield  {title} {\enquote {\bibinfo {title}
  {On the nonorthogonality problem},}\ }\href {\doibase
  http://dx.doi.org/10.1016/S0065-3276(08)60339-1} {\bibfield  {journal}
  {\bibinfo  {journal} {Adv. Quantum Chem.}\ }\textbf {\bibinfo {volume} {5}},\
  \bibinfo {pages} {185} (\bibinfo {year} {1970})}\BibitemShut {NoStop}%
\bibitem [{\citenamefont {Liu}\ and\ \citenamefont
  {Li}(1997)}]{liu-a-method-1997}%
  \BibitemOpen
  \bibfield  {author} {\bibinfo {author} {\bibfnamefont {Wenjian}\ \bibnamefont
  {Liu}}\ and\ \bibinfo {author} {\bibfnamefont {Lemin}\ \bibnamefont {Li}},\
  }\bibfield  {title} {\enquote {\bibinfo {title} {A method for population and
  bonding analyses in calculations with extended basis sets},}\ }\href
  {\doibase 10.1007/BF02341693} {\bibfield  {journal} {\bibinfo  {journal}
  {Theor. Chim. Acta}\ }\textbf {\bibinfo {volume} {95}},\ \bibinfo {pages}
  {81} (\bibinfo {year} {1997})}\BibitemShut {NoStop}%
\bibitem [{\citenamefont {Mohr}\ \emph {et~al.}(2014)\citenamefont {Mohr},
  \citenamefont {Ratcliff}, \citenamefont {Boulanger}, \citenamefont
  {Genovese}, \citenamefont {Caliste}, \citenamefont {Deutsch},\ and\
  \citenamefont {Goedecker}}]{mohr-daubechies-2014}%
  \BibitemOpen
  \bibfield  {author} {\bibinfo {author} {\bibfnamefont {Stephan}\ \bibnamefont
  {Mohr}}, \bibinfo {author} {\bibfnamefont {Laura~E.}\ \bibnamefont
  {Ratcliff}}, \bibinfo {author} {\bibfnamefont {Paul}\ \bibnamefont
  {Boulanger}}, \bibinfo {author} {\bibfnamefont {Luigi}\ \bibnamefont
  {Genovese}}, \bibinfo {author} {\bibfnamefont {Damien}\ \bibnamefont
  {Caliste}}, \bibinfo {author} {\bibfnamefont {Thierry}\ \bibnamefont
  {Deutsch}}, \ and\ \bibinfo {author} {\bibfnamefont {Stefan}\ \bibnamefont
  {Goedecker}},\ }\bibfield  {title} {\enquote {\bibinfo {title} {Daubechies
  wavelets for linear scaling density functional theory},}\ }\href {\doibase
  http://dx.doi.org/10.1063/1.4871876} {\bibfield  {journal} {\bibinfo
  {journal} {J. Chem. Phys.}\ }\textbf {\bibinfo {volume} {140}},\ \bibinfo
  {eid} {204110} (\bibinfo {year} {2014})}\BibitemShut {NoStop}%
\bibitem [{\citenamefont {Mohr}\ \emph {et~al.}(2015)\citenamefont {Mohr},
  \citenamefont {Ratcliff}, \citenamefont {Genovese}, \citenamefont {Caliste},
  \citenamefont {Boulanger}, \citenamefont {Goedecker},\ and\ \citenamefont
  {Deutsch}}]{mohr-accurate-2015}%
  \BibitemOpen
  \bibfield  {author} {\bibinfo {author} {\bibfnamefont {Stephan}\ \bibnamefont
  {Mohr}}, \bibinfo {author} {\bibfnamefont {Laura~E.}\ \bibnamefont
  {Ratcliff}}, \bibinfo {author} {\bibfnamefont {Luigi}\ \bibnamefont
  {Genovese}}, \bibinfo {author} {\bibfnamefont {Damien}\ \bibnamefont
  {Caliste}}, \bibinfo {author} {\bibfnamefont {Paul}\ \bibnamefont
  {Boulanger}}, \bibinfo {author} {\bibfnamefont {Stefan}\ \bibnamefont
  {Goedecker}}, \ and\ \bibinfo {author} {\bibfnamefont {Thierry}\ \bibnamefont
  {Deutsch}},\ }\bibfield  {title} {\enquote {\bibinfo {title} {Accurate and
  efficient linear scaling dft calculations with universal applicability},}\
  }\href {\doibase 10.1039/C5CP00437C} {\bibfield  {journal} {\bibinfo
  {journal} {Phys. Chem. Chem. Phys.}\ }\textbf {\bibinfo {volume} {17}},\
  \bibinfo {pages} {31360} (\bibinfo {year} {2015})}\BibitemShut {NoStop}%
\bibitem [{\citenamefont {Haynes}\ \emph {et~al.}(2006)\citenamefont {Haynes},
  \citenamefont {Skylaris}, \citenamefont {Mostofi},\ and\ \citenamefont
  {Payne}}]{haynes-onetep-2006}%
  \BibitemOpen
  \bibfield  {author} {\bibinfo {author} {\bibfnamefont {Peter~D.}\
  \bibnamefont {Haynes}}, \bibinfo {author} {\bibfnamefont {Chris-Kriton}\
  \bibnamefont {Skylaris}}, \bibinfo {author} {\bibfnamefont {Arash~A.}\
  \bibnamefont {Mostofi}}, \ and\ \bibinfo {author} {\bibfnamefont {Mike~C.}\
  \bibnamefont {Payne}},\ }\bibfield  {title} {\enquote {\bibinfo {title}
  {Onetep: linear-scaling density-functional theory with local orbitals and
  plane waves},}\ }\href {\doibase 10.1002/pssb.200541457} {\bibfield
  {journal} {\bibinfo  {journal} {Phys. Status Solidi B}\ }\textbf {\bibinfo
  {volume} {243}},\ \bibinfo {pages} {2489} (\bibinfo {year}
  {2006})}\BibitemShut {NoStop}%
\bibitem [{\citenamefont {Mostofi}\ \emph {et~al.}(2007)\citenamefont
  {Mostofi}, \citenamefont {Haynes}, \citenamefont {Skylaris},\ and\
  \citenamefont {Payne}}]{mostofi-onetep-2007}%
  \BibitemOpen
  \bibfield  {author} {\bibinfo {author} {\bibfnamefont {A.~A.}\ \bibnamefont
  {Mostofi}}, \bibinfo {author} {\bibfnamefont {P.~D.}\ \bibnamefont {Haynes}},
  \bibinfo {author} {\bibfnamefont {C.~K.}\ \bibnamefont {Skylaris}}, \ and\
  \bibinfo {author} {\bibfnamefont {M.~C.}\ \bibnamefont {Payne}},\ }\bibfield
  {title} {\enquote {\bibinfo {title} {Onetep: linear-scaling
  density-functional theory with plane-waves},}\ }\href {\doibase
  10.1080/08927020600932801} {\bibfield  {journal} {\bibinfo  {journal} {Mol.
  Simul.}\ }\textbf {\bibinfo {volume} {33}},\ \bibinfo {pages} {551} (\bibinfo
  {year} {2007})},\ \Eprint
  {http://arxiv.org/abs/http://dx.doi.org/10.1080/08927020600932801}
  {http://dx.doi.org/10.1080/08927020600932801} \BibitemShut {NoStop}%
\bibitem [{\citenamefont {Skylaris}\ \emph {et~al.}(2008)\citenamefont
  {Skylaris}, \citenamefont {Haynes}, \citenamefont {Mostofi},\ and\
  \citenamefont {Payne}}]{skylaris-recent-2008}%
  \BibitemOpen
  \bibfield  {author} {\bibinfo {author} {\bibfnamefont {Chris-Kriton}\
  \bibnamefont {Skylaris}}, \bibinfo {author} {\bibfnamefont {Peter~D}\
  \bibnamefont {Haynes}}, \bibinfo {author} {\bibfnamefont {Arash~A}\
  \bibnamefont {Mostofi}}, \ and\ \bibinfo {author} {\bibfnamefont {Mike~C}\
  \bibnamefont {Payne}},\ }\bibfield  {title} {\enquote {\bibinfo {title}
  {Recent progress in linear-scaling density functional calculations with plane
  waves and pseudopotentials: the onetep code},}\ }\href
  {http://stacks.iop.org/0953-8984/20/i=6/a=064209} {\bibfield  {journal}
  {\bibinfo  {journal} {J. Phys.: Condens. Matter}\ }\textbf {\bibinfo {volume}
  {20}},\ \bibinfo {pages} {064209} (\bibinfo {year} {2008})}\BibitemShut
  {NoStop}%
\bibitem [{\citenamefont {Bowler}\ \emph {et~al.}(2000)\citenamefont {Bowler},
  \citenamefont {Bush},\ and\ \citenamefont {Gillan}}]{bowler-practical-2000}%
  \BibitemOpen
  \bibfield  {author} {\bibinfo {author} {\bibfnamefont {D.~R.}\ \bibnamefont
  {Bowler}}, \bibinfo {author} {\bibfnamefont {I.~J.}\ \bibnamefont {Bush}}, \
  and\ \bibinfo {author} {\bibfnamefont {M.~J.}\ \bibnamefont {Gillan}},\
  }\bibfield  {title} {\enquote {\bibinfo {title} {Practical methods for ab
  initio calculations on thousands of atoms},}\ }\href {\doibase
  10.1002/(SICI)1097-461X(2000)77:5<831::AID-QUA5>3.0.CO;2-G} {\bibfield
  {journal} {\bibinfo  {journal} {Int. J. Quantum Chem.}\ }\textbf {\bibinfo
  {volume} {77}},\ \bibinfo {pages} {831} (\bibinfo {year} {2000})}\BibitemShut
  {NoStop}%
\bibitem [{\citenamefont {Bowler}\ and\ \citenamefont
  {Miyazaki}(2010)}]{bowler-an-overview-2010}%
  \BibitemOpen
  \bibfield  {author} {\bibinfo {author} {\bibfnamefont {D~R}\ \bibnamefont
  {Bowler}}\ and\ \bibinfo {author} {\bibfnamefont {T}~\bibnamefont
  {Miyazaki}},\ }\bibfield  {title} {\enquote {\bibinfo {title} {{Calculations
  for millions of atoms with density functional theory: linear scaling shows
  its potential.}}}\ }\href {\doibase 10.1088/0953-8984/22/7/074207} {\bibfield
   {journal} {\bibinfo  {journal} {J. Phys.: Condens. Matter}\ }\textbf
  {\bibinfo {volume} {22}},\ \bibinfo {pages} {074207} (\bibinfo {year}
  {2010})}\BibitemShut {NoStop}%
\bibitem [{\citenamefont {VandeVondele}\ \emph {et~al.}(2005)\citenamefont
  {VandeVondele}, \citenamefont {Krack}, \citenamefont {Mohamed}, \citenamefont
  {Parrinello}, \citenamefont {Chassaing},\ and\ \citenamefont
  {Hutter}}]{vandevondele-quickstep-2005}%
  \BibitemOpen
  \bibfield  {author} {\bibinfo {author} {\bibfnamefont {Joost}\ \bibnamefont
  {VandeVondele}}, \bibinfo {author} {\bibfnamefont {Matthias}\ \bibnamefont
  {Krack}}, \bibinfo {author} {\bibfnamefont {Fawzi}\ \bibnamefont {Mohamed}},
  \bibinfo {author} {\bibfnamefont {Michele}\ \bibnamefont {Parrinello}},
  \bibinfo {author} {\bibfnamefont {Thomas}\ \bibnamefont {Chassaing}}, \ and\
  \bibinfo {author} {\bibfnamefont {J\"{u}rg}\ \bibnamefont {Hutter}},\
  }\bibfield  {title} {\enquote {\bibinfo {title} {{Quickstep: Fast and
  accurate density functional calculations using a mixed Gaussian and plane
  waves approach}},}\ }\href {\doibase 10.1016/j.cpc.2004.12.014} {\bibfield
  {journal} {\bibinfo  {journal} {Comput. Phys. Commun.}\ }\textbf {\bibinfo
  {volume} {167}},\ \bibinfo {pages} {103} (\bibinfo {year}
  {2005})}\BibitemShut {NoStop}%
\bibitem [{\citenamefont {Soler}\ \emph {et~al.}(2002)\citenamefont {Soler},
  \citenamefont {Artacho}, \citenamefont {Gale}, \citenamefont {Garc\'ia},
  \citenamefont {Junquera}, \citenamefont {Ordej\'on},\ and\ \citenamefont
  {S\'anchez-Portal}}]{soler-the_siesta-2002}%
  \BibitemOpen
  \bibfield  {author} {\bibinfo {author} {\bibfnamefont {Jos\'e~M}\
  \bibnamefont {Soler}}, \bibinfo {author} {\bibfnamefont {Emilio}\
  \bibnamefont {Artacho}}, \bibinfo {author} {\bibfnamefont {Julian~D}\
  \bibnamefont {Gale}}, \bibinfo {author} {\bibfnamefont {Alberto}\
  \bibnamefont {Garc\'ia}}, \bibinfo {author} {\bibfnamefont {Javier}\
  \bibnamefont {Junquera}}, \bibinfo {author} {\bibfnamefont {Pablo}\
  \bibnamefont {Ordej\'on}}, \ and\ \bibinfo {author} {\bibfnamefont {Daniel}\
  \bibnamefont {S\'anchez-Portal}},\ }\bibfield  {title} {\enquote {\bibinfo
  {title} {The siesta method for ab initio order- n materials simulation},}\
  }\href {http://stacks.iop.org/0953-8984/14/i=11/a=302} {\bibfield  {journal}
  {\bibinfo  {journal} {J. Phys.: Condens. Matter}\ }\textbf {\bibinfo {volume}
  {14}},\ \bibinfo {pages} {2745} (\bibinfo {year} {2002})}\BibitemShut
  {NoStop}%
\bibitem [{\citenamefont {Artacho}\ \emph {et~al.}(2008)\citenamefont
  {Artacho}, \citenamefont {Anglada}, \citenamefont {Di\'eguez}, \citenamefont
  {Gale}, \citenamefont {Garc\'ia}, \citenamefont {Junquera}, \citenamefont
  {Martin}, \citenamefont {Ordej\'on}, \citenamefont {Pruneda}, \citenamefont
  {S\'anchez-Portal},\ and\ \citenamefont {Soler}}]{artacho-the_siesta-2008}%
  \BibitemOpen
  \bibfield  {author} {\bibinfo {author} {\bibfnamefont {Emilio}\ \bibnamefont
  {Artacho}}, \bibinfo {author} {\bibfnamefont {E}~\bibnamefont {Anglada}},
  \bibinfo {author} {\bibfnamefont {O}~\bibnamefont {Di\'eguez}}, \bibinfo
  {author} {\bibfnamefont {J~D}\ \bibnamefont {Gale}}, \bibinfo {author}
  {\bibfnamefont {A}~\bibnamefont {Garc\'ia}}, \bibinfo {author} {\bibfnamefont
  {J}~\bibnamefont {Junquera}}, \bibinfo {author} {\bibfnamefont {R~M}\
  \bibnamefont {Martin}}, \bibinfo {author} {\bibfnamefont {P}~\bibnamefont
  {Ordej\'on}}, \bibinfo {author} {\bibfnamefont {J~M}\ \bibnamefont
  {Pruneda}}, \bibinfo {author} {\bibfnamefont {D}~\bibnamefont
  {S\'anchez-Portal}}, \ and\ \bibinfo {author} {\bibfnamefont {J~M}\
  \bibnamefont {Soler}},\ }\bibfield  {title} {\enquote {\bibinfo {title} {The
  siesta method; developments and applicability},}\ }\href
  {http://stacks.iop.org/0953-8984/20/i=6/a=064208} {\bibfield  {journal}
  {\bibinfo  {journal} {J. Phys.: Condens. Matter}\ }\textbf {\bibinfo {volume}
  {20}},\ \bibinfo {pages} {064208} (\bibinfo {year} {2008})}\BibitemShut
  {NoStop}%
\bibitem [{\citenamefont {Daubechies}(1992)}]{daubechies-ten-1992}%
  \BibitemOpen
  \bibfield  {author} {\bibinfo {author} {\bibfnamefont {Ingrid}\ \bibnamefont
  {Daubechies}},\ }\href@noop {} {\emph {\bibinfo {title} {{Ten lectures on
  wavelets}}}}\ (\bibinfo  {publisher} {Society for Industrial and Applied
  Mathematics},\ \bibinfo {address} {Philadelphia},\ \bibinfo {year}
  {1992})\BibitemShut {NoStop}%
\bibitem [{\citenamefont {Hartwigsen}\ \emph {et~al.}(1998)\citenamefont
  {Hartwigsen}, \citenamefont {Goedecker},\ and\ \citenamefont
  {Hutter}}]{hartwigsen-relativistic-1998}%
  \BibitemOpen
  \bibfield  {author} {\bibinfo {author} {\bibfnamefont {C.}~\bibnamefont
  {Hartwigsen}}, \bibinfo {author} {\bibfnamefont {S.}~\bibnamefont
  {Goedecker}}, \ and\ \bibinfo {author} {\bibfnamefont {J.}~\bibnamefont
  {Hutter}},\ }\bibfield  {title} {\enquote {\bibinfo {title} {Relativistic
  separable dual-space gaussian pseudopotentials from h to rn},}\ }\href
  {\doibase 10.1103/PhysRevB.58.3641} {\bibfield  {journal} {\bibinfo
  {journal} {Phys. Rev. B}\ }\textbf {\bibinfo {volume} {58}},\ \bibinfo
  {pages} {3641} (\bibinfo {year} {1998})}\BibitemShut {NoStop}%
\bibitem [{\citenamefont {Willand}\ \emph {et~al.}(2013)\citenamefont
  {Willand}, \citenamefont {Kvashnin}, \citenamefont {Genovese}, \citenamefont
  {V{\'{a}}zquez-Mayagoitia}, \citenamefont {Deb}, \citenamefont {Sadeghi},
  \citenamefont {Deutsch},\ and\ \citenamefont
  {Goedecker}}]{willand-norm-conserving-2013}%
  \BibitemOpen
  \bibfield  {author} {\bibinfo {author} {\bibfnamefont {Alex}\ \bibnamefont
  {Willand}}, \bibinfo {author} {\bibfnamefont {Yaroslav~O}\ \bibnamefont
  {Kvashnin}}, \bibinfo {author} {\bibfnamefont {Luigi}\ \bibnamefont
  {Genovese}}, \bibinfo {author} {\bibfnamefont {{\'{A}}lvaro}\ \bibnamefont
  {V{\'{a}}zquez-Mayagoitia}}, \bibinfo {author} {\bibfnamefont
  {Arpan~Krishna}\ \bibnamefont {Deb}}, \bibinfo {author} {\bibfnamefont {Ali}\
  \bibnamefont {Sadeghi}}, \bibinfo {author} {\bibfnamefont {Thierry}\
  \bibnamefont {Deutsch}}, \ and\ \bibinfo {author} {\bibfnamefont {Stefan}\
  \bibnamefont {Goedecker}},\ }\bibfield  {title} {\enquote {\bibinfo {title}
  {{Norm-conserving pseudopotentials with chemical accuracy compared to
  all-electron calculations}},}\ }\href {\doibase
  http://dx.doi.org/10.1063/1.4793260} {\bibfield  {journal} {\bibinfo
  {journal} {J. Chem. Phys.}\ }\textbf {\bibinfo {volume} {138}},\ \bibinfo
  {pages} {104109} (\bibinfo {year} {2013})}\BibitemShut {NoStop}%
\bibitem [{\citenamefont {Perdew}\ \emph {et~al.}(1996)\citenamefont {Perdew},
  \citenamefont {Burke},\ and\ \citenamefont
  {Ernzerhof}}]{perdew-generalized-1996}%
  \BibitemOpen
  \bibfield  {author} {\bibinfo {author} {\bibfnamefont {John~P.}\ \bibnamefont
  {Perdew}}, \bibinfo {author} {\bibfnamefont {Kieron}\ \bibnamefont {Burke}},
  \ and\ \bibinfo {author} {\bibfnamefont {Matthias}\ \bibnamefont
  {Ernzerhof}},\ }\bibfield  {title} {\enquote {\bibinfo {title} {Generalized
  gradient approximation made simple},}\ }\href {\doibase
  10.1103/PhysRevLett.77.3865} {\bibfield  {journal} {\bibinfo  {journal}
  {Phys. Rev. Lett.}\ }\textbf {\bibinfo {volume} {77}},\ \bibinfo {pages}
  {3865} (\bibinfo {year} {1996})}\BibitemShut {NoStop}%
\bibitem [{\citenamefont {Genovese}\ \emph {et~al.}(2008)\citenamefont
  {Genovese}, \citenamefont {Neelov}, \citenamefont {Goedecker}, \citenamefont
  {Deutsch}, \citenamefont {Ghasemi}, \citenamefont {Willand}, \citenamefont
  {Caliste}, \citenamefont {Zilberberg}, \citenamefont {Rayson}, \citenamefont
  {Bergman},\ and\ \citenamefont {Schneider}}]{genovese-daubechies-2008}%
  \BibitemOpen
  \bibfield  {author} {\bibinfo {author} {\bibfnamefont {Luigi}\ \bibnamefont
  {Genovese}}, \bibinfo {author} {\bibfnamefont {Alexey}\ \bibnamefont
  {Neelov}}, \bibinfo {author} {\bibfnamefont {Stefan}\ \bibnamefont
  {Goedecker}}, \bibinfo {author} {\bibfnamefont {Thierry}\ \bibnamefont
  {Deutsch}}, \bibinfo {author} {\bibfnamefont {Seyed~Alireza}\ \bibnamefont
  {Ghasemi}}, \bibinfo {author} {\bibfnamefont {Alexander}\ \bibnamefont
  {Willand}}, \bibinfo {author} {\bibfnamefont {Damien}\ \bibnamefont
  {Caliste}}, \bibinfo {author} {\bibfnamefont {Oded}\ \bibnamefont
  {Zilberberg}}, \bibinfo {author} {\bibfnamefont {Mark}\ \bibnamefont
  {Rayson}}, \bibinfo {author} {\bibfnamefont {Anders}\ \bibnamefont
  {Bergman}}, \ and\ \bibinfo {author} {\bibfnamefont {Reinhold}\ \bibnamefont
  {Schneider}},\ }\bibfield  {title} {\enquote {\bibinfo {title} {{Daubechies
  wavelets as a basis set for density functional pseudopotential
  calculations.}}}\ }\href {\doibase 10.1063/1.2949547} {\bibfield  {journal}
  {\bibinfo  {journal} {J. Chem. Phys.}\ }\textbf {\bibinfo {volume} {129}},\
  \bibinfo {pages} {014109} (\bibinfo {year} {2008})}\BibitemShut {NoStop}%
\bibitem [{\citenamefont {Jensen}(2013)}]{jensen-atomic-2013}%
  \BibitemOpen
  \bibfield  {author} {\bibinfo {author} {\bibfnamefont {Frank}\ \bibnamefont
  {Jensen}},\ }\bibfield  {title} {\enquote {\bibinfo {title} {Atomic orbital
  basis sets},}\ }\href {\doibase 10.1002/wcms.1123} {\bibfield  {journal}
  {\bibinfo  {journal} {Wiley Interdiscip. Rev.-Comput. Mol. Sci.}\ }\textbf
  {\bibinfo {volume} {3}},\ \bibinfo {pages} {273} (\bibinfo {year}
  {2013})}\BibitemShut {NoStop}%
\bibitem [{\citenamefont {Case}\ \emph {et~al.}(2005)\citenamefont {Case},
  \citenamefont {Cheatham}, \citenamefont {Darden}, \citenamefont {Gohlke},
  \citenamefont {Luo}, \citenamefont {Merz}, \citenamefont {Onufriev},
  \citenamefont {Simmerling}, \citenamefont {Wang},\ and\ \citenamefont
  {Woods}}]{case-the-2005}%
  \BibitemOpen
  \bibfield  {author} {\bibinfo {author} {\bibfnamefont {David~A.}\
  \bibnamefont {Case}}, \bibinfo {author} {\bibfnamefont {Thomas~E.}\
  \bibnamefont {Cheatham}}, \bibinfo {author} {\bibfnamefont {Tom}\
  \bibnamefont {Darden}}, \bibinfo {author} {\bibfnamefont {Holger}\
  \bibnamefont {Gohlke}}, \bibinfo {author} {\bibfnamefont {Ray}\ \bibnamefont
  {Luo}}, \bibinfo {author} {\bibfnamefont {Kenneth~M.}\ \bibnamefont {Merz}},
  \bibinfo {author} {\bibfnamefont {Alexey}\ \bibnamefont {Onufriev}}, \bibinfo
  {author} {\bibfnamefont {Carlos}\ \bibnamefont {Simmerling}}, \bibinfo
  {author} {\bibfnamefont {Bing}\ \bibnamefont {Wang}}, \ and\ \bibinfo
  {author} {\bibfnamefont {Robert~J.}\ \bibnamefont {Woods}},\ }\bibfield
  {title} {\enquote {\bibinfo {title} {{The Amber biomolecular simulation
  programs}},}\ }\href {\doibase 10.1002/jcc.20290} {\bibfield  {journal}
  {\bibinfo  {journal} {J. Comput. Chem.}\ }\textbf {\bibinfo {volume} {26}},\
  \bibinfo {pages} {1668} (\bibinfo {year} {2005})},\ \Eprint
  {http://arxiv.org/abs/NIHMS150003} {arXiv:NIHMS150003} \BibitemShut {NoStop}%
\bibitem [{\citenamefont {Case}\ \emph {et~al.}()\citenamefont {Case},
  \citenamefont {Darden}, \citenamefont {Cheatham}, \citenamefont {Simmerling},
  \citenamefont {Wang}, \citenamefont {Duke}, \citenamefont {Luo},
  \citenamefont {Crowley}, \citenamefont {Walker}, \citenamefont {Zhang},
  \citenamefont {Merz}, \citenamefont {Wang}, \citenamefont {Hayik},
  \citenamefont {Roitberg}, \citenamefont {Seabra}, \citenamefont
  {Kolossv\'{a}ry}, \citenamefont {Wong}, \citenamefont {Paesani},
  \citenamefont {Vanicek}, \citenamefont {Wu}, \citenamefont {Brozell},
  \citenamefont {Steinbrecher}, \citenamefont {Gohlke}, \citenamefont {Yang},
  \citenamefont {Tan}, \citenamefont {Mongan}, \citenamefont {Hornak},
  \citenamefont {Cui}, \citenamefont {Mathews}, \citenamefont {Seetin},
  \citenamefont {Sagui}, \citenamefont {Babin},\ and\ \citenamefont
  {Kollman}}]{Amber11}%
  \BibitemOpen
  \bibfield  {author} {\bibinfo {author} {\bibfnamefont {David~A.}\
  \bibnamefont {Case}}, \bibinfo {author} {\bibfnamefont {T.~A.}\ \bibnamefont
  {Darden}}, \bibinfo {author} {\bibfnamefont {T.~E.}\ \bibnamefont
  {Cheatham}}, \bibinfo {author} {\bibfnamefont {Carlos~L.}\ \bibnamefont
  {Simmerling}}, \bibinfo {author} {\bibfnamefont {J.}~\bibnamefont {Wang}},
  \bibinfo {author} {\bibfnamefont {Robert~E.}\ \bibnamefont {Duke}}, \bibinfo
  {author} {\bibfnamefont {Ray}\ \bibnamefont {Luo}}, \bibinfo {author}
  {\bibfnamefont {Michael}\ \bibnamefont {Crowley}}, \bibinfo {author}
  {\bibfnamefont {Ross~C.}\ \bibnamefont {Walker}}, \bibinfo {author}
  {\bibfnamefont {W.}~\bibnamefont {Zhang}}, \bibinfo {author} {\bibfnamefont
  {K.~M.}\ \bibnamefont {Merz}}, \bibinfo {author} {\bibfnamefont
  {B.}~\bibnamefont {Wang}}, \bibinfo {author} {\bibfnamefont {S.}~\bibnamefont
  {Hayik}}, \bibinfo {author} {\bibfnamefont {Adrian}\ \bibnamefont
  {Roitberg}}, \bibinfo {author} {\bibfnamefont {Gustavo}\ \bibnamefont
  {Seabra}}, \bibinfo {author} {\bibfnamefont {I.}~\bibnamefont
  {Kolossv\'{a}ry}}, \bibinfo {author} {\bibfnamefont {K.~F.}\ \bibnamefont
  {Wong}}, \bibinfo {author} {\bibfnamefont {F.}~\bibnamefont {Paesani}},
  \bibinfo {author} {\bibfnamefont {J.}~\bibnamefont {Vanicek}}, \bibinfo
  {author} {\bibfnamefont {X.}~\bibnamefont {Wu}}, \bibinfo {author}
  {\bibfnamefont {Scott~R.}\ \bibnamefont {Brozell}}, \bibinfo {author}
  {\bibfnamefont {Tom}\ \bibnamefont {Steinbrecher}}, \bibinfo {author}
  {\bibfnamefont {Holger}\ \bibnamefont {Gohlke}}, \bibinfo {author}
  {\bibfnamefont {L.}~\bibnamefont {Yang}}, \bibinfo {author} {\bibfnamefont
  {C.}~\bibnamefont {Tan}}, \bibinfo {author} {\bibfnamefont {J.}~\bibnamefont
  {Mongan}}, \bibinfo {author} {\bibfnamefont {V.}~\bibnamefont {Hornak}},
  \bibinfo {author} {\bibfnamefont {G.}~\bibnamefont {Cui}}, \bibinfo {author}
  {\bibfnamefont {D.~H.}\ \bibnamefont {Mathews}}, \bibinfo {author}
  {\bibfnamefont {M.~G.}\ \bibnamefont {Seetin}}, \bibinfo {author}
  {\bibfnamefont {C.}~\bibnamefont {Sagui}}, \bibinfo {author} {\bibfnamefont
  {V.}~\bibnamefont {Babin}}, \ and\ \bibinfo {author} {\bibfnamefont
  {Peter~A.}\ \bibnamefont {Kollman}},\ }\href@noop {} {\emph {\bibinfo {title}
  {{Amber 11}}}},\ \bibinfo {organization} {University of California, San
  Francisco}\BibitemShut {NoStop}%
\bibitem [{\citenamefont {Hornak}\ \emph {et~al.}(2006)\citenamefont {Hornak},
  \citenamefont {Abel}, \citenamefont {Okur}, \citenamefont {Strockbine},
  \citenamefont {Roitberg},\ and\ \citenamefont
  {Simmerling}}]{hornak-comparison-2006}%
  \BibitemOpen
  \bibfield  {author} {\bibinfo {author} {\bibfnamefont {Viktor}\ \bibnamefont
  {Hornak}}, \bibinfo {author} {\bibfnamefont {Robert}\ \bibnamefont {Abel}},
  \bibinfo {author} {\bibfnamefont {Asim}\ \bibnamefont {Okur}}, \bibinfo
  {author} {\bibfnamefont {Bentley}\ \bibnamefont {Strockbine}}, \bibinfo
  {author} {\bibfnamefont {Adrian}\ \bibnamefont {Roitberg}}, \ and\ \bibinfo
  {author} {\bibfnamefont {Carlos}\ \bibnamefont {Simmerling}},\ }\bibfield
  {title} {\enquote {\bibinfo {title} {Comparison of multiple amber force
  fields and development of improved protein backbone parameters},}\ }\href
  {\doibase 10.1002/prot.21123} {\bibfield  {journal} {\bibinfo  {journal}
  {Proteins Struct. Funct. Bioinf.}\ }\textbf {\bibinfo {volume} {65}},\
  \bibinfo {pages} {712} (\bibinfo {year} {2006})}\BibitemShut {NoStop}%
\bibitem [{\citenamefont {Momma}\ and\ \citenamefont
  {Izumi}(2011)}]{momma-VESTA-2011}%
  \BibitemOpen
  \bibfield  {author} {\bibinfo {author} {\bibfnamefont {Koichi}\ \bibnamefont
  {Momma}}\ and\ \bibinfo {author} {\bibfnamefont {Fujio}\ \bibnamefont
  {Izumi}},\ }\bibfield  {title} {\enquote {\bibinfo {title} {{VESTA 3 for
  three-dimensional visualization of crystal, volumetric and morphology
  data}},}\ }\href {\doibase 10.1107/S0021889811038970} {\bibfield  {journal}
  {\bibinfo  {journal} {J. Appl. Crystallogr.}\ }\textbf {\bibinfo {volume}
  {44}},\ \bibinfo {pages} {1272} (\bibinfo {year} {2011})}\BibitemShut
  {NoStop}%
\bibitem [{\citenamefont {Stone}(1981)}]{stone-distributed-1981}%
  \BibitemOpen
  \bibfield  {author} {\bibinfo {author} {\bibfnamefont {A.~J.}\ \bibnamefont
  {Stone}},\ }\bibfield  {title} {\enquote {\bibinfo {title} {{Distributed
  multipole analysis, or how to describe a molecular charge distribution}},}\
  }\href {\doibase 10.1016/0009-2614(81)85452-8} {\bibfield  {journal}
  {\bibinfo  {journal} {Chem. Phys. Lett.}\ }\textbf {\bibinfo {volume} {83}},\
  \bibinfo {pages} {233} (\bibinfo {year} {1981})}\BibitemShut {NoStop}%
\bibitem [{\citenamefont {Sokalski}\ and\ \citenamefont
  {Poirier}(1983)}]{sokalski-cumulative-1983}%
  \BibitemOpen
  \bibfield  {author} {\bibinfo {author} {\bibfnamefont {W.~Andrzej}\
  \bibnamefont {Sokalski}}\ and\ \bibinfo {author} {\bibfnamefont {R.~A.}\
  \bibnamefont {Poirier}},\ }\bibfield  {title} {\enquote {\bibinfo {title}
  {{Cumulative atomic multipole representation of the molecular charge
  distribution and its basis set dependence}},}\ }\href {\doibase
  10.1016/0009-2614(83)80208-5} {\bibfield  {journal} {\bibinfo  {journal}
  {Chem. Phys. Lett.}\ }\textbf {\bibinfo {volume} {98}},\ \bibinfo {pages}
  {86} (\bibinfo {year} {1983})}\BibitemShut {NoStop}%
\bibitem [{\citenamefont {Williams}(1988)}]{williams-representation-1988}%
  \BibitemOpen
  \bibfield  {author} {\bibinfo {author} {\bibfnamefont {Donald~E.}\
  \bibnamefont {Williams}},\ }\bibfield  {title} {\enquote {\bibinfo {title}
  {{Representation of the molecular electrostatic potential by atomic multipole
  and bond dipole models}},}\ }\href {\doibase 10.1002/jcc.540090705}
  {\bibfield  {journal} {\bibinfo  {journal} {J. Comput. Chem.}\ }\textbf
  {\bibinfo {volume} {9}},\ \bibinfo {pages} {745} (\bibinfo {year}
  {1988})}\BibitemShut {NoStop}%
\bibitem [{\citenamefont {Sokalski}\ \emph {et~al.}(1992)\citenamefont
  {Sokalski}, \citenamefont {Shibata}, \citenamefont {Rein},\ and\
  \citenamefont {Ornstein}}]{sokalski-cumulative-1992}%
  \BibitemOpen
  \bibfield  {author} {\bibinfo {author} {\bibfnamefont {W.A.}\ \bibnamefont
  {Sokalski}}, \bibinfo {author} {\bibfnamefont {M.}~\bibnamefont {Shibata}},
  \bibinfo {author} {\bibfnamefont {R.}~\bibnamefont {Rein}}, \ and\ \bibinfo
  {author} {\bibfnamefont {R.L.}\ \bibnamefont {Ornstein}},\ }\bibfield
  {title} {\enquote {\bibinfo {title} {{Cumulative atomic multipole moments
  complement any atomic charge model to obtain more accurate electrostatic
  properties}},}\ }\href {\doibase 10.1002/jcc.540130713} {\bibfield  {journal}
  {\bibinfo  {journal} {J. Comput. Chem.}\ }\textbf {\bibinfo {volume} {13}},\
  \bibinfo {pages} {883} (\bibinfo {year} {1992})}\BibitemShut {NoStop}%
\bibitem [{\citenamefont {Whitehead}\ \emph {et~al.}(2003)\citenamefont
  {Whitehead}, \citenamefont {Breneman}, \citenamefont {Sukumar},\ and\
  \citenamefont {Ryan}}]{whitehead-transferable-2003}%
  \BibitemOpen
  \bibfield  {author} {\bibinfo {author} {\bibfnamefont {C.~E.}\ \bibnamefont
  {Whitehead}}, \bibinfo {author} {\bibfnamefont {C.~M.}\ \bibnamefont
  {Breneman}}, \bibinfo {author} {\bibfnamefont {N.}~\bibnamefont {Sukumar}}, \
  and\ \bibinfo {author} {\bibfnamefont {M.~D.}\ \bibnamefont {Ryan}},\
  }\bibfield  {title} {\enquote {\bibinfo {title} {{Transferable atom
  equivalent multicentered multipole expansion method}},}\ }\href {\doibase
  10.1002/jcc.10240} {\bibfield  {journal} {\bibinfo  {journal} {J. Comput.
  Chem.}\ }\textbf {\bibinfo {volume} {24}},\ \bibinfo {pages} {512} (\bibinfo
  {year} {2003})}\BibitemShut {NoStop}%
\bibitem [{\citenamefont {Day}\ \emph {et~al.}(2005)\citenamefont {Day},
  \citenamefont {Motherwell},\ and\ \citenamefont {Jones}}]{day-beyond-2005}%
  \BibitemOpen
  \bibfield  {author} {\bibinfo {author} {\bibfnamefont {Graeme~M.}\
  \bibnamefont {Day}}, \bibinfo {author} {\bibfnamefont {W.~D.~Sam}\
  \bibnamefont {Motherwell}}, \ and\ \bibinfo {author} {\bibfnamefont
  {William}\ \bibnamefont {Jones}},\ }\bibfield  {title} {\enquote {\bibinfo
  {title} {{Beyond the Isotropic Atom Model in Crystal Structure Prediction of
  Rigid Molecules: Atomic Multipoles versus Point Charges}},}\ }\href {\doibase
  10.1021/cg049651n} {\bibfield  {journal} {\bibinfo  {journal} {Cryst. Growth
  Des.}\ }\textbf {\bibinfo {volume} {5}},\ \bibinfo {pages} {1023} (\bibinfo
  {year} {2005})}\BibitemShut {NoStop}%
\bibitem [{\citenamefont {Plattner}\ and\ \citenamefont
  {Meuwly}(2009)}]{plattner-higher-2009}%
  \BibitemOpen
  \bibfield  {author} {\bibinfo {author} {\bibfnamefont {Nuria}\ \bibnamefont
  {Plattner}}\ and\ \bibinfo {author} {\bibfnamefont {Markus}\ \bibnamefont
  {Meuwly}},\ }\bibfield  {title} {\enquote {\bibinfo {title} {{Higher order
  multipole moments for molecular dynamics simulations}},}\ }\href {\doibase
  10.1007/s00894-009-0465-6} {\bibfield  {journal} {\bibinfo  {journal} {J.
  Mol. Model.}\ }\textbf {\bibinfo {volume} {15}},\ \bibinfo {pages} {687}
  (\bibinfo {year} {2009})}\BibitemShut {NoStop}%
\bibitem [{\citenamefont {Kramer}\ \emph {et~al.}(2013)\citenamefont {Kramer},
  \citenamefont {Bereau}, \citenamefont {Spinn}, \citenamefont {Liedl},
  \citenamefont {Gedeck},\ and\ \citenamefont {Meuwly}}]{kramer-deriving-2013}%
  \BibitemOpen
  \bibfield  {author} {\bibinfo {author} {\bibfnamefont {Christian}\
  \bibnamefont {Kramer}}, \bibinfo {author} {\bibfnamefont {Tristan}\
  \bibnamefont {Bereau}}, \bibinfo {author} {\bibfnamefont {Alexander}\
  \bibnamefont {Spinn}}, \bibinfo {author} {\bibfnamefont {Klaus~R.}\
  \bibnamefont {Liedl}}, \bibinfo {author} {\bibfnamefont {Peter}\ \bibnamefont
  {Gedeck}}, \ and\ \bibinfo {author} {\bibfnamefont {Markus}\ \bibnamefont
  {Meuwly}},\ }\bibfield  {title} {\enquote {\bibinfo {title} {{Deriving static
  atomic multipoles from the electrostatic potential}},}\ }\href {\doibase
  10.1021/ci400548w} {\bibfield  {journal} {\bibinfo  {journal} {J. Chem. Inf.
  Model.}\ }\textbf {\bibinfo {volume} {53}},\ \bibinfo {pages} {3410}
  (\bibinfo {year} {2013})}\BibitemShut {NoStop}%
\bibitem [{\citenamefont {Mohr}\ \emph {et~al.}(to be submitted)\citenamefont
  {Mohr}, \citenamefont {Masella}, \citenamefont {Ratcliff},\ and\
  \citenamefont {Genovese}}]{mohr-fragments2-2017}%
  \BibitemOpen
  \bibfield  {author} {\bibinfo {author} {\bibfnamefont {Stephan}\ \bibnamefont
  {Mohr}}, \bibinfo {author} {\bibfnamefont {Michel}\ \bibnamefont {Masella}},
  \bibinfo {author} {\bibfnamefont {Laura~E.}\ \bibnamefont {Ratcliff}}, \ and\
  \bibinfo {author} {\bibfnamefont {Luigi}\ \bibnamefont {Genovese}},\
  }\bibfield  {title} {\enquote {\bibinfo {title} {Complexity reduction in
  large quantum systems: Reliable electrostatic embedding for multiscale
  approaches via optimized minimal basis functions},}\ }\href@noop {} {\
  (\bibinfo {year} {to be submitted})}\BibitemShut {NoStop}%
\bibitem [{\citenamefont {Rico}\ \emph {et~al.}(2013)\citenamefont {Rico},
  \citenamefont {L\'opez}, \citenamefont {Ema},\ and\ \citenamefont
  {Ram\'irez}}]{RSHshift}%
  \BibitemOpen
  \bibfield  {author} {\bibinfo {author} {\bibfnamefont {Jaime~Fern\'andez}\
  \bibnamefont {Rico}}, \bibinfo {author} {\bibfnamefont {Rafael}\ \bibnamefont
  {L\'opez}}, \bibinfo {author} {\bibfnamefont {Ignacio}\ \bibnamefont {Ema}},
  \ and\ \bibinfo {author} {\bibfnamefont {Guillermo}\ \bibnamefont
  {Ram\'irez}},\ }\bibfield  {title} {\enquote {\bibinfo {title} {Translation
  of real solid spherical harmonics},}\ }\href {\doibase 10.1002/qua.24356}
  {\bibfield  {journal} {\bibinfo  {journal} {Int. J. Quantum Chem.}\ }\textbf
  {\bibinfo {volume} {113}},\ \bibinfo {pages} {1544} (\bibinfo {year}
  {2013})}\BibitemShut {NoStop}%
\end{thebibliography}%



%

\end{document}